\documentclass{emulateapj}
\usepackage[usenames]{color}

\newcommand{\kms}{\mbox{km s$^{-1}$}}

\newcommand{\mhi}{$M_{\rm HI}$}%

\newcommand{\bbf}{}

\slugcomment{Draft version December 1, 2008}

\shorttitle{Disentangling the circumnuclear environs of Centaurus A I. }
\shortauthors{D. Espada et al.}

\begin{document}

\title{Disentangling the circumnuclear environs of Centaurus A: \\
 I. High resolution molecular gas imaging}

\author{D. Espada \altaffilmark{1,2,3}, S. Matsushita \altaffilmark{1}, A. Peck \altaffilmark{4,5},
C. Henkel \altaffilmark{6}, 
D. Iono \altaffilmark{7}, 
F. P. Israel \altaffilmark{8},
S. Muller \altaffilmark{1},
G. Petitpas\altaffilmark{5}, 
Y. Pihlstr{\"o}m \altaffilmark{9}, 
G. B. Taylor \altaffilmark{9}, and
Dinh-V-Trung \altaffilmark{1,10}
}

\altaffiltext{1}{Academia Sinica, Institute of Astronomy and Astrophysics, P.O. Box 23-141, Taipei 106, Taiwan. } 
\altaffiltext{2}{Harvard-Smithsonian Center for Astrophysics, 60 Garden St., Cambridge, MA 02138; despada@cfa.harvard.edu}
\altaffiltext{3}{Instituto de Astrof{\'i}sica de Andaluc{\'i}a - CSIC, Apdo. 3004, 18080 Granada, Spain}
\altaffiltext{4}{Joint ALMA Office, Av. El Golf 40, Piso 18, Las Condes, Santiago, Chile }
\altaffiltext{5}{Harvard-Smithsonian Center for Astrophysics, Submillimeter Array, 645 North A`ohoku Place, Hilo, HI 96720}
\altaffiltext{6}{Max-Planck-Institut f{\"u}r Radioastronomie, Auf dem H{\"u}gel 69, 53121 Bonn, Germany} 
\altaffiltext{7}{National Astronomical Observatory of Japan, 2-21-1 Osawa, Mitaka, Tokyo 181-8588, Japan} 
\altaffiltext{8}{Sterrewacht Leiden, Niels-Bohr-Weg 2, 2300 CA Leiden, The Netherlands}
\altaffiltext{9}{Department of Physics and Astronomy, MSC07 4220, University of New Mexico, Albuquerque, NM 87131.  Also Adjunct Astronomers at the National Radio Astronomy Observatory}
\altaffiltext{10}{On leave from Institute of Physics, Vietnamese Academy of Science \& Technology, 10, Daotan, BaDinh, Hanoi, Vietnam}

\begin{abstract}
{\bbf  We present high resolution images of the $^{12}$CO(2--1)
  emission in the central 1\arcmin\ (1~kpc) of NGC~5128 (Centaurus A),
  observed using the Submillimeter Array.  We elucidate for the first
  time the distribution and kinematics of the molecular gas in this
  region with a resolution of 6\farcs0 $\times$ 2\farcs4 (100~pc
  $\times$ 40~pc). 
 We spatially resolve the
  circumnuclear molecular gas in the inner 24$\arcsec$ $\times$
  12$\arcsec$ (400 pc $\times$ 200 pc), which is elongated along a
  position angle P.A.\ $\simeq$ $155\arcdeg$ and perpendicular to the
  radio/X-ray jet. The SE and NW components of the circumnuclear gas are connected
  to molecular gas found at larger radii.  This gas appears as two parallel
  filaments at P.A.\ = $120\arcdeg$, which are coextensive with the
  long sides of the 3 kiloparsec parallelogram shape of the previously observed dust continuum, as well as
  ionized and pure rotational H$_2$ lines.  Spatial and kinematical
  asymmetries are apparent in both the circumnuclear and outer gas,
  suggesting non-coplanar and/or non-circular motions. We extend to
  inner radii ($r$ $<$ 200~pc) previously studied warped disk models
  built to reproduce the central parallelogram-shaped structure.
Adopting the warped disk model we would confirm a gap in emission
between the radii $r$ = 200 -- 800~pc (12\arcsec -- 50\arcsec), as has
been suggested previously.
Although this model explains this prominent feature,
however, our $\rm ^{12}CO(2--1)$ observations show relevant deviations from this model.  Namely, the physical
connection between the circumnuclear gas and that at larger radii,
brighter SE and NW sides on the parallelogram-shaped feature, and an
outer curvature of its long sides. Overall it resembles more closely an
S-shaped morphology, a trend that is also found in other molecular species.
Hence, we explore qualitatively the possible contribution of a weak
bi-symmetric potential which would naturally explain these
peculiarities. }
\end{abstract}


\section{Introduction}
\label{introduction}

Radio galaxies are radio-loud active galaxies, usually of elliptical type. Their large scale synchrotron jets are presumably powered by the accretion of gas onto supermassive black holes, which are fuelled by reservoirs of neutral and ionized gas in the host galaxy.  Many of these radio galaxies possess dust lanes containing large amounts of the different components of the interstellar medium (e.g. \citealt{2002ApJS..139..411A}).  Cold gas traced by CO rotational lines is seen in some of these galaxies, with double-horned line profiles suggesting rotating disks in their nuclear region (e.g. \citealt{2003ASPC..290..525L,2003ASPC..290..529L}).  The origin of the cold gas in these galaxies is still debated but it may be the result of mergers with smaller gas rich galaxies.

Understanding the distribution, kinematics and physical conditions of the nuclear gas in radio galaxies is essential to constrain the properties of the central engines of the most powerful active galactic nuclei (AGN). Questions that need to be addressed include: What are the mechanisms that drive the gas from kiloparsec to parsec scales?  What are the gas concentrations in the nuclear region? Are the collimated and relativistic jets linked somehow to the gas seen at large scales?  Powerful radio sources are rare in the local Universe, and thus the lack of high resolution observations has prevented the systematic study of the properties of the molecular gas in their nuclear region.

 Centaurus~A (Cen~A, NGC~5128) is by far the nearest and best studied radio galaxy and thus plays an important role in our understanding of this major class of active galaxies. At its distance of $D\sim3.42$~Mpc (\citealt{2007ApJ...654..186F}), $1\arcmin$ is approximately only 1~kpc, thus provides a unique opportunity to study the nuclear molecular gas in extraordinary detail. Cen~A is often considered to be the prototypical Fanaroff-Riley (FR) class I low-luminosity radio galaxy hosting a type 2 AGN according to unified schemes \citep{1993ARA&A..31..473A}.   Table~\ref{tbl-1} contains a summary of the characteristics of this galaxy from recent literature.  For a detailed and comprehensive review, refer to \citet{1998A&ARv...8..237I}.


At optical wavelengths the most characteristic feature of the galaxy,
other than its typical elliptical appearance, is the warped and almost
edge-on dust lane located along the minor axis of the galaxy.  This
dust lane contains a large amount of atomic, molecular and ionized gas
\citep{1998A&ARv...8..237I}.  Atomic hydrogen emission follows the
dust lane and extends further out, to a radius of at least 7~kpc,
showing asymmetries that indicate that the outer parts are not well
ordered \citep{1990AJ.....99.1781V,1994ApJ...423L.101S}.  The central
region of Cen~A (from 3~kpc to sub-pc scale) is very complex.  The
most prominent components are described as follows:

i) A \emph{disk-like structure associated with the dust lane} ($\sim$
5\arcmin , or 5~kpc) as observed in H$\alpha$
(e.g. \citealt{1992ApJ...387..503N}), NIR \citep{1993ApJ...412..550Q},
submillimeter continuum
\citep[e.g.][]{1993MNRAS.260..844H,2002ApJ...565..131L}, CO lines
\citep[e.g.][]{1987ApJ...322L..73P,1990ApJ...365..522E,1993A&A...270L..13R,2001AA...371..865L}
and mid-IR continuum
\citep[e.g.][]{1999A&A...341..667M,2006ApJ...645.1092Q}. {\bbf A
  remarkably symmetric structure is found within this disk in the
  inner 3 kpc.  This feature is distributed along the dust lane at a
  position angle of 120\arcdeg , and is usually described as an {\it
    S-shaped} or a {\it parallelogram structure} (a rhomboid-like
  feature with long and short sides of length 130\arcsec\ and
  50\arcsec , respectively).  The former description has been
  used for example for ISOCAM mid-IR
  \citep{1999A&A...341..667M}, JCMT submm
  (e.g. \citealt{2002ApJ...565..131L}) and recent Spitzer/IRS H$_2$
  S(0) 28 $\mu$m observations \citep{2008MNRAS.384.1469Q}. The latter
  description is used in Spitzer/IRAC \citep{2006ApJ...645.1092Q} and
  Spitzer/IRS mid-IR observations \citep{2008MNRAS.384.1469Q} of the
  continuum, ionized gas emission such as [\ion{Ne}{2}] 12.8
  $\mu$m, [\ion{S}{3}] 33.4$\mu$m and [\ion{Si}{2}] 34.8 $\mu$m and
  PAHs (polyaromatic hydrocarbons). However, the two nomenclatures
  refer to the same structure and imply that emission in some species
  are more prominent in the SE and NW of the long sides of the
  parallelogram structure.
This parallelogram structure has been well modelled by
\citet{2006ApJ...645.1092Q} using a non-coplanar model with circular
motions, a slightly modified version of the warped disk model based on
Spitzer/IRAC data, but originally obtained from previous CO, near-IR
\citep{1992ApJ...391..121Q,1993ApJ...412..550Q} and H$\alpha$
observations
\citep{1986PhDT.......143B,1987MNRAS.228..595B,1992ApJ...387..503N}. Alternatively,
it has been proposed that the observed features could be due to a
ring-like structure within a bar
\citep{1999A&A...341..667M,2000ApJ...528..276M}, which would partially
explain the S-shaped morphology of some species. }

ii) A \emph{rapidly rotating circumnuclear region} within the central few 100\,pc (6\arcsec\ -- 25\arcsec) of the galaxy, as suggested from broad molecular lines (mainly CO, e.g. \citealt{1990A&A...227..342I,1991A&A...245L..13I,1992A&A...265..487I,1993A&A...270L..13R}).  Although previous single dish CO observations lack the angular resolution to resolve the circumnuclear region and to separate it from the outer disk of Cen~A, the spectra toward the center show a broad plateau between 300~km~s$^{-1} < $V$ <$ 800~km~s$^{-1}$ that was interpreted to be a circumnuclear molecular disk with an extent of a few 100~pc \citep{1991A&A...245L..13I,1993A&A...270L..13R}.  This component is also indicated by mid-IR/submm emission (e.g. \citealt{1993MNRAS.260..844H}; \citealt{1999A&A...341..667M}) and Pa$\alpha$ emission  \citep{1998ApJ...499L.143S,2000ApJ...528..276M}, but remained undetected in H$\alpha$ mainly due to extinction  (e.g. \citealt{1992ApJ...387..503N}).

iii) A \emph{nuclear disk} ($\sim$ 30~pc, or 2\arcsec) containing ionized and molecular gas presumably feeding a nuclear massive object \citep[e.g.][]{2001ApJ...549..915M,2006A&A...448..921M}.  While  ionized gas species show non-rotational motions and are likely related to the jets, the molecular hydrogen as traced by H$_2$(J=1--0) S(1) seems to be well distributed within a disk-like structure \citep{2007ApJ...671.1329N}.  Cen~A hosts a compact sub-pc sized nuclear radio continuum source exhibiting notable intensity variations at radio, infrared and X-ray wavelengths, presumably powered by accretion events \citep{1998A&ARv...8..237I,2008A&A...483..741I}.

iv) \emph{Absorption lines} toward the central continuum source detected in \ion{H}{1} (e.g. \citealt{1970ApJ...161L...9R,1983ApJ...264L..37V,2002ApJ...564..696S}) and molecular lines (e.g. \citealt{1990A&A...227..342I,1997A&A...324...51W,1999ApJ...516..769E}), which provide the possibility to study the properties of the atomic and molecular clouds along the line of sight.

\setcounter{footnote}{0}

Since Cen~A is located in the southern hemisphere (Declination  $\simeq -43\arcdeg $), high resolution mapping of CO lines was not feasible until the advent of the Submillimeter Array (SMA\footnote{The Submillimeter Array is a joint project between the Smithsonian Astrophysical Observatory and the Academia Sinica Institute of Astronomy and Astrophysics, and is funded by the Smithsonian Institution and the Academia Sinica.}; \citealt{2004ApJ...616L...1H}).  Here we report the findings of our $^{12}$CO(2--1) observations with the SMA of its nuclear region with a resolution of 6\farcs0 $\times$ 2\farcs4 (100 $\times$ 40 pc).  We use these high resolution interferometric observations to reveal the morphology and kinematics of the circumnuclear and outer molecular gas, as well as to elucidate the connection between these two molecular gas components.  Further goals are to study the connection between molecular gas, dust, ionized gas and jets, and the evaluation of a model that best reproduces the observed features.

The paper is organized as follows:  We introduce our SMA $^{12}$CO(2--1) observations and data reduction in \S~\ref{observationReduction}. In \S~\ref{result} we focus on the identification of the different components and study their physical properties.  Possible models that can reproduce the CO emission are evaluated in \S~\ref{model}.  In \S~\ref{previousWork} we compare our data with those from other wavelengths using the wealth of information from previous studies in the literature.  We discuss our results in \S~\ref{discussion} and finally summarize our conclusions in \S~\ref{conclusion}.  In a companion paper \citep{2008ApJ...000..000E}, we study the properties of the SMA $^{12}$CO(2--1) absorption spectrum toward the compact bright continuum as well as VLBA \ion{H}{1} absorption lines against both the nucleus and the nuclear jet.

\section{SMA $^{12}$CO(2--1) Observations and Data Reduction}
\label{observationReduction}

Cen~A was observed at 1.3~mm using the SMA with 7 antennas on April 5, 2006.  In  Table~\ref{tbl-2}, we summarize the main parameters of our interferometric observations.  The digital correlator was configured with 3072 channels (2~GHz bandwidth), resulting in a velocity resolution of about 1~\kms.  The receivers were tuned to the redshifted $^{12}$CO(2--1) ($\nu_{\rm rest}$ = 230.538~GHz) emission line in the upper sideband (USB), using $V_{\rm LSR}$ = 550~\kms . Note that velocities are expressed throughout this paper with respect to the LSR using the radio convention. 
This setting allowed us to simultaneously obtain the $^{13}$CO(2--1) ($\nu_{\rm rest}$ = 220.397~GHz) and C$^{18}$O(2--1) ($\nu_{\rm rest}$ = 219.560~GHz) lines in the lower sideband (LSB).  
We used R.A.\ = $13^{\rm h}25^{\rm m}27\fs6$ and Dec.\ = $-43\arcdeg 01 \arcmin 09\farcs0$ (J2000) as our phase center, which is $0\farcs2$ offset from the AGN position R.A.\ = $13^{\rm h}25^{\rm m}27\fs615$ and Dec.\ = $-43\arcdeg 01 \arcmin 08\farcs805$ \citep[J2000;][]{1998AJ....116..516M}.  
The maximum elevation of the source at the SMA site is $\simeq 27\arcdeg$, forcing us to observe only under very stable atmospheric conditions, with zenith opacities of typically $\tau_{225}\sim0.10$.  In order to have a beam shape as close to circular as possible we used a compact configuration with longer N--S baselines.  Unprojected baseline lengths spanned 16 -- 70~m, corresponding to 6 -- 30~m projected baselines.

The editing and calibration of the data were done with the SMA-adapted MIR software\footnote{MIR is a software package to reduce SMA data based on the package originally developed by Nick Scoville at Caltech.  See \url{http://cfa-www.harvard.edu/$\sim$cqi/mircook.html}}.  Callisto and 3C273 were used for passband calibration.  An initial gain calibration was performed using J1316--3338, which is at an angular distance of $9\arcdeg$ from the target.  The continuum emission toward Cen~A was found to consist of an unresolved source at its center.  The gain calibration was thus refined using the averaged line-free channels in Cen~A itself.  3C273 was used as absolute flux calibrator.  Although this source is variable, it is frequently monitored at the SMA, and we adopted the value of $11.4\pm0.8$~Jy measured at 1~mm on 13 April, 2006.  Overall, we estimate the absolute flux uncertainties on the order of 20\%.

The imaging of the $^{12}$CO(2--1) line was conducted in MIRIAD \citep{1995ASPC...77..433S}.  A careful subtraction of the continuum was done using line-free channels with the task \verb!UVLIN!.  Since the continuum emission is so strong in this source, we paid special attention that continuum subtraction was done correctly by checking for possible artifacts in the line-free channels.
The data were \verb!CLEAN!ed with uniform weighting.  This weighting was used to minimize the sidelobe level rather than the noise.  Minimizing the sidelobe level is especially important in the N--S direction for channels close to the systemic velocity, where confusion by emission from different components, as well as the absorption features, can be a major problem in the cleaning process.  The field of view (F.O.V.) is characterized by a Half Power Beam Width (HPBW) of the primary beam of an SMA antenna of $52\arcsec$ (0.9~kpc), and the resolution by the final synthesized beam of $2\farcs4 \times 6\farcs0$ ($40 \times 100$~pc) with a major axis P.A.\ = $0.2\arcdeg$.  The mean rms noise level is 64~mJy beam$^{-1}$ for the uniformly weighted 10~\kms\ \verb!CLEAN!ed channel maps.  The task \verb!MOMENT! was used to calculate the $^{12}$CO(2--1) integrated flux density distribution and the intensity-weighted velocity field distribution (clipped at 2 times the rms noise).

\section{Results}
\label{result}

{\color{red}




}

The 1.3~mm continuum source was characterized by a total flux of $5.9\pm1.0$~Jy and was found to be unresolved.  The $^{12}$CO(2--1) line was detected in emission and absorption, $^{13}$CO(2--1) only in absorption and C$^{18}$O(2--1) was not detected in either emission or absorption.

In Figure~\ref{fig-compara-spectra} we compare the $^{12}$CO(2--1)
profile of \citet{1992A&A...265..487I} using the Swedish-ESO
Submillimetre Telescope (SEST) with our SMA $^{12}$CO(2--1) profile
obtained by convolving the data to the $23\arcsec$ HPBW of the SEST
and extracting the profile at the phase center. The $^{12}$CO(2--1)
emission line shows a broad plateau and a narrow line, in agreement
with previous single-dish observations
(e.g. \citealt{1991A&A...245L..13I}). The total integrated intensity
of the $^{12}$CO(2--1) line as seen from our convolved maps is
$1790\pm70$~Jy~\kms, recovering around 85\% of the total flux measured
using the SEST (1890~Jy~\kms; \citealt{1992A&A...265..487I}). This is
well within the absolute flux calibration uncertainties. We do not
miss a large fraction of the emission as a result of the lack of
short-spacings, but we find systematicly lower flux densities in the
narrow line component (velocity range close to the systemic velocity).
This component might therefore correspond to extended emission that is
being filtered out.  In contrast, the broad plateau should correspond
to a quite compact spatial component, since we recover all the flux.

In Figure~\ref{fig1} we show the channel maps covering the velocity interval from 320 to 810~km~s$^{-1}$ in 10~\kms\ bins.  Figures~\ref{fig2} and \ref{fig3} show the $^{12}$CO(2--1) integrated intensity (primary beam corrected) and velocity field maps, respectively.  Position--Velocity (P--V) diagrams along several cuts are shown in Figure~\ref{fig-pv}.  Figure~\ref{fig-pv} a) and b) show the P--V diagrams along P.A.\ = $120\arcdeg$, for two different offsets in declination at $-10\arcsec$ and $+10\arcsec$ with respect to the center of the map. Figure~\ref{fig-pv} c) and d) show the P--V diagrams along the P.A.\ = $135\arcdeg$ and $155\arcdeg$, respectively. 

From the $^{12}$CO(2--1) emission distributions and kinematics, we are able to distinguish between three components: {\bbf 1) the broad plateau}, corresponding to emission within the plotted ellipse around the center of the galaxy, in the form of a rotating circumnuclear structure (as outlined in \S~\ref{introduction}, ii); {\bbf 2) the narrow line component}, along the two parallel lines plotted from SE to NW with a P.A.\ = $120\arcdeg$ and offset from the center at Dec. = $-10\arcsec$ and $+10\arcsec$, associated to the dust lane (gas at larger radii from the center, \S~\ref{introduction}, i); and {\bbf 3) the absorption features }toward the unresolved compact continuum component located in the galactic center (\S~\ref{introduction}, iv), which will be described in a forthcoming paper \citep{2008ApJ...000..000E}.
With our resolution, we are not able to resolve the nuclear disk  (\S~\ref{introduction}, iii).

In the following, we describe the observed broad plateau and narrow line components in more detail.  
A summary of the derived parameters for the $^{12}$CO(2--1) emission for each component is shown in  Table~\ref{tbl-3}, including the peak flux density, mean velocity, velocity width and the total flux density ($S_{\rm CO(2-1)}$).

\subsection{{\bbf Broad Plateau}: Rapidly rotating circumnuclear gas}
\label{subsect:circumnucleardisk}


           
The $^{12}$CO(2--1) channel maps (Figure~\ref{fig1}) clearly show highly blueshifted ($\sim$340 -- 420~km~s$^{-1}$) emission in  the SE side of the AGN location and redshifted ($\sim$660 -- 760~km~s$^{-1}$) emission on the NW side.  This velocity range corresponds to the broad component of the spectrum that has a full width to zero power of  $\Delta V$ $\simeq$ 420 $\pm$ 14 \kms, spanning from 340 to 760~\kms\ (Figure~\ref{fig-compara-spectra}).  The total integrated intensity and velocity field maps (Figure~\ref{fig2} and \ref{fig3}) show that this broad line plateau is located within the overlaid ellipse. The major and minor axes of this ellipse have lengths of $24\arcsec \times 12\arcsec$ (400~pc $\times$ 200~pc in linear scale without taking into account any projection effect) with a P.A.\ $=155\arcdeg$, centered on the AGN. {\bbf  The P--V diagram along the major axis of this ellipse (Figure~\ref{fig-pv} d) shows a gradient of 19.2~\kms~arcsec$^{-1}$ (or 1.2~\kms~pc$^{-1}$), which can be explained either by a disk or ring-like structure.} 
Given that this broad line component is located exclusively in the inner 24\arcsec\  (400 pc), it must arise from an extended circumnuclear region. Although
           the emission is barely resolved in the N--S direction
           due to the elongated beam, the assumption of a 
           (to first order) circular disk would imply that we are 
           viewing the molecular component at an inclination angle 
           of at least $i$ = 70$^{\circ}$.

\subsection{Narrow Line: Gas at larger radii associated with the dust lane}
\label{subsect:dustlane}

The channel maps (Figure~\ref{fig1}) also show emission, distributed
along two parallel filaments ($\pm10\arcsec$ north and south of the
nucleus) with P.A. $\simeq$ 120$\arcdeg$, from the SE edge of the
F.O.V.\ at velocities $V$ = 430~km~s$^{-1}$, up to the NW edge with
$V$ = 650~km~s$^{-1}$. These features correspond to the narrow
($\sim100$~\kms) line component previously observed in single-dish CO
spectra (Figure~\ref{fig-compara-spectra}; see also
e.g. \citealt{1990A&A...227..342I}). The P.A.\ of the circumnuclear
gas and that of this component differ by $\Delta$P.A. $\simeq$
35$\arcdeg$.  The two parallel features can be seen more clearly in
the integrated intensity map (Figure~\ref{fig2}). 
The velocity gradient of both
of these two parallel features (Figures~\ref{fig3} and \ref{fig-pv} a,
b) is 3.2~\kms~arcsec$^{-1}$, which is 6 times smaller than that of
the circumnuclear gas.  The two (northern and southern) parallel
features are quite similar in distribution and kinematics and must be
somehow related to each other.  In a zeroth-order approximation, both
features would be in agreement with a highly inclined rotating
ring located at a much larger radius than that of the
circumnuclear gas, for which we only see the central $\sim$
$60\arcsec$ due to the restricted F.O.V.\ of our observations, at
230~GHz.  However, the molecular gas is likely characterized by
non-coplanar and/or non-circular motions, as explained in
\S~\ref{subsect:noncirc}.

\subsection{Warped and/or Non-Circular Motions}
\label{subsect:noncirc}

{\color{red}









                   }

 We confirm from our $^{12}$CO(2--1) data that the molecular gas is
 likely non-coplanar and/or contains non-circular motions. Evidence is
 summarized as follows: {\bbf i) Large scale warp or non-circular
   motions in the gas at large radii}: if emission arises from a
 coplanar (ring-like) feature, the connecting lines between the two
 lumps in each channel map (Figure~\ref{fig1}, for example in channels
 570 -- 630 \kms) should be perpendicular to the major axis at P.A.\ =
 $120\arcdeg$. However we find an apparent misalignment. This can also
 be seen in the velocity field (Figure~\ref{fig3}). Moreover, in
 channels within the 450 -- 500 \kms\ range in Figure~\ref{fig1} the
 northern lump seems to be superposed on the southern filament. {\bbf
   ii) Brighter emission on the SE and NW sides in the two parallel
   filaments:} on the receding side there is more emission in the
 northern filament than in the southern one, while the opposite is
 found on the approaching side (Figure~\ref{fig2}). {\bbf iii) Outer
   curvature:} the emission does not exactly follow the drawn lines
 (Figure~\ref{fig2}), but there is a curvature as we go farther from
 the centre. {\bbf iv) Connections between circumnuclear and outer
   gas}: are found in the range $V$ = 480 -- 570~\kms\ in the channel
 maps (Figure~\ref{fig1}). The smooth connection of the gas kinematics
 in the SE between both components can be seen in the P--V diagram at
 angular offsets $8\arcsec - 12\arcsec$ along P.A. = 135$\arcdeg$ and
 155$\arcdeg$ (Figures~\ref{fig-pv} c and d).  {\bbf v) Warp and/or
   non-circular motions in the circumnuclear gas: } although the
 high-velocity parts ($|V - V_{sys}| > $150~\kms, where $V_{sys}$ is
 the systemic velocity) are quite symmetric and well centered
 kinematically, there are obvious deviations from co-planar and/or
 circular motions in the circumnuclear region at velocities close to
 $V_{sys}$.  Whereas the receding part ($V$ = 620 -- 680~\kms) is
 quite similar to a disk-like feature, the approaching side ($V$ = 420
 -- 480~\kms) is not, with more emission located in the eastern side
 of the ellipse than in the region closer to the AGN.

{\bbf Given the complexity of the circumnuclear region, we plot in Figure~\ref{fig-spectrumComposition} a grid of emission line profiles. We discern the double-peaked profiles in  the intersection regions between the circumnuclear and outer gas. We find velocity dispersions of about $\sigma$ = 50 -- 60~\kms\ (corresponding to a full width half maximum FWHM = 110 -- 130~\kms) in the circumnuclear gas, which may be due to larger velocity gradients. The velocity dispersion in the molecular gas associated with the dust lane is 5 times smaller.
The concentration of the circumnuclear gas peaks close to these intersection points. This double-peaked distribution in the circumnuclear gas may be similar to the twin peak feature often seen in the central region of barred galaxies \citep[e.g.][]{1992ApJ...395L..79K,1999ApJ...511..157K}.  
A caveat to the interpretation of the twin peak feature is the presence of a strong absorption line. Indeed, emission is seen to weaken in the inner regions from $V$ = 500 to 600~\kms, but this can be explained only in part by the absorption features located on the red-shifted side ($V$ = 544 --  615 \kms) of the $V_{sys}$ of Cen~A \citep{2008ApJ...000..000E}.  Therefore the twin peak interpretation is plausible. Moreover, the connection between the circumnuclear and outer gas found in the P--V diagram in the SE (Figures~\ref{fig-pv} c and d) reinforce this idea.

Note that the superposition of emission arising from the circumnuclear
       gas and that from the gas at larger radii
       causes a biased first moment. This is because the central value of the fit shifts as the relative prominence of the lines changes. For that reason we do not derive any values of velocities or widths from moment data. Nevertheless, Figure~\ref{fig3} is relevant since it illustrates the clear trend of rotation apparent in the circumnuclear gas.}


\subsection{Molecular Gas Mass}
\label{sub:moleculargasmass}

{\color{red}

      }

{\bbf We separate the emission of the circumnuclear molecular gas from
  the molecular gas at larger radii in order to estimate their
  respective masses.  We report in Table~\ref{tbl-3} the calculated
  molecular gas masses ($M_{\rm H_{2}}$) of both components.  For the
  circumnuclear molecular gas mass we take into account a conversion
  factor between integrated CO intensity and H$_2$ column density $X$
  $= N_{H_2} /I _{CO}$ = 0.4 $\times$
  10$^{20}$~cm$^{-2}$~(K~km~s$^{-1}$)$^{-1}$, which is a typical value
  observed in the nuclear regions of galaxies
  \citep{1988ApJ...325..389M,1995ApJ...448L..97W,1996A&A...305..421M,2001A&A...365..571W}. For
  the gas at larger radii, we adopt instead $X$ $= 2.0 \times
  10^{20}$~cm$^{-2}$~(K~km~s$^{-1}$)$^{-1}$, which is a mean value for
  $|b|$ $<$ 5\arcdeg\ in the Galaxy
  \citep{2001ApJ...547..792D}. The conversion factor $X$ may be
  uncertain by a factor of two even if Cen~A were a spiral galaxy. Since
  the assumption of Cen~A being a spiral obviously fails, a systematic error to $X$ may be present
  which cannot be evaluated due to a lack of statistically relevant
  observational data.
We also  use a line ratio between $^{12}$CO(1--0) and (2--1) lines over the whole molecular disk of Cen~A of $T_{\rm CO(2-1)} / T_{\rm CO(1-0)} \simeq 0.9$ \citep{1993A&A...270L..13R}, and accounting for the
       $\nu^2$ factor when converting flux densities
       to line temperatures.
       The integrated intensity of the circumnuclear gas 
       is reduced by only about 2\% of its total flux by the absorption line ($\int S~dV = 25.6$~$\pm$~$0.5$~Jy~beam$^{-1}$~\kms, \citealt{2008ApJ...000..000E}), and 
       therefore does not significantly affect our mass estimate.
       
}

We find the molecular gas masses of the circumnuclear gas and gas at
larger radii to be $M_{\rm H_{2}}^{c}$ =
$(0.6\pm0.1)\times10^7$~M$_\odot$ and $M_{\rm H_{2}}^{o}$
=$(5.9\pm0.3)\times10^7$~M$_\odot$, respectively, indicating that the
outer component in our maps is about an order of magnitude as massive
as the circumnuclear gas.  We have detected a total molecular gas mass
of $M_{\rm H_{2}} \simeq (6.5 \pm 0.3) \times 10^7$~M$_\odot$. Note
that errors do not include the uncertainties mainly arising from the
$X$ factor discussed above. The gas masses derived here can be
re-scaled by a different conversion factor if a more appropriate value
is found. 
Note that our maps (primary beam corrected) detect only a part
of the narrow line component, so the true mass is larger than our
estimate.  From previously reported single-dish observations, the total
$^{12}$CO(1--0) luminosity is $L_{\rm CO} = 10.8 \times
10^7$~K~\kms~pc$^2$ \citep{1990ApJ...363..451E}, which yields a
molecular gas mass of M$_{\rm H_{2}}^{total}$ $\simeq$ $2.3 \times
10^8$~M$_{\odot}$, using $X = 2.0 \times
10^{20}$~cm$^{-2}$~(K~km~s$^{-1}$)$^{-1}$. Since the integrated flux
density from the circumnuclear gas is just a small fraction of the
total ($\sim$ 1/6), the use of this conversion factor is a good
estimate to derive the total gas mass. We therefore have in the region observed with the SMA (52\arcsec\ HPBW) about 30 -- 40 \% of
the total molecular gas mass.


%


The total gas mass in the inner 1\arcmin\ (1 kpc) is 
$M_{\rm gas}$ $\simeq (9.0 \pm 0.4) \times 10^7$~M$_\odot$  
using $M_{\rm gas} = 1.36 \times M_{\rm H_{2}}$, where the factor 1.36 accounts for elements other than hydrogen \citep{1973asqu.book.....A,1999ApJS..124..403S}. From that amount, there is 
$M_{\rm gas}^c$ $\simeq (0.8 \pm 0.1) \times 10^7$~M$_\odot$
in the inner projected distance $r$ $<$ 200 pc (12\arcsec). 
The surface density within $r$ = 200~pc is 
$\Sigma_{\rm gas, 200~pc} \simeq 65$~M$_\odot$~pc$^{-2}$ 
assuming a disk-like feature with inclination $i$ = $70\arcdeg$.
The dynamical mass within the circumnuclear gas can be calculated as $M_{\rm dyn, 200~pc} = r~(V / \sin i)^2 / G = 4.6 \times 10^9$~M$_{\odot}$, as obtained from the maximum rotational velocity $V \simeq 200$~km~s$^{-1}$ corrected for an inclination of $i$ = $70\arcdeg$.  Thus the molecular gas in Cen~A constitutes {\bbf less than 2\% of the dynamical total mass within the radius $r$ = 200 pc}.

\section{Models: Warped Thin Disk, Weak Bar and Combined Models}
\label{model}

{\color{red}

}


There is evidence that the gas and dust are relatively well settled in the inner 5\arcmin\ (5 kpc) (e.g. \citealt{1992ApJ...387..503N, 1993A&A...270L..13R,2006ApJ...645.1092Q}). 
  \citet{1992ApJ...387..503N} and \citet{1992ApJ...391..121Q} have shown, with H$\alpha$ and CO respectively, that co-planar motions are not in agreement with the observations. \citet{2006ApJ...645.1092Q} present a slightly revised version of a warped disk model using circular orbits, based on previous CO, near-IR  \citep{1992ApJ...391..121Q,1993ApJ...412..550Q} and H$\alpha$ observations \citep{1986PhDT.......143B,1987MNRAS.228..595B,1992ApJ...387..503N}, which succesfully  reproduce the observed parallelogram structure \citep{2006ApJ...645.1092Q}. 


 {\bbf Another proposed interpretation is an embedded large scale
   ($\sim$ 5\arcmin, or 5 kpc) stellar bar structure
   \citep[][mid-IR]{1999A&A...341..667M}, but analysis of the detailed
   applicability of this model has not been performed. One point
   against this interpretation is the fact that the total molecular
   gas mass accounts only for about 2\% of the dynamical gas mass
   (\S~\ref{sub:moleculargasmass}) so the molecular gas motions are
   expected to be mainly dominated by the stellar component (and dark
   matter), but no non-axisymmetry of the stellar component has been
   seen or inferred from kinematic observations or mass models of the
   central stellar region. \citet{2006ApJ...645.1092Q} have also
   searched for evidence of a large-scale stellar bar in the isophote
   shapes 3.6 $\mu$m Spitzer/IRAC images, which is not affected by
   extinction, and obtained negative results.  On the other hand,
   although a strong bar model can probably be discarded due to the lack of
   a prominent non-axisymmetric stellar component, it would still be
   difficult to detect a weak non-axisymmetric stellar component (such
   as that of a putative weak 'bar'), because the disk is nearly
   edge-on, masked by the dust lane, and embedded in the bright
   spheroidal component.  Evidence for a nuclear bar has been
   suggested for example by \citet{2001ApJ...549..915M}, who find that
   VLT/ISAAC spectra along different P.A. in the central 25\arcsec\ of
   H$_2$(J=1--0) S(1) ($\sim$ 500 pc) is probably characterized by
   non-circular motions.  }

The lack of observations of the gas with sufficient spatial resolution from 100~pc to kiloparsec scale as well as kinematical information, in addition to  high extinction  ($A_{\rm V} = 27 \pm 5$~mag, \citealt{1990A&A...227..342I}),  has further limited our knowledge on boundary conditions that are essential to model the circumnuclear region of Cen~A.


{\bbf 
 In this section, we investigate a comparison of our $^{12}$CO(2--1) data  (\S~\ref{result}) with the revised warped disk model in \citet[][]{2006ApJ...645.1092Q}.  
 Encouraged by the observed deviations with respect to this model (\S~\ref{subsect:warp}), we propose that non-circular motions via a weak bi-symmetric potential may partly govern the distribution and kinematics of the molecular gas  (\S~\ref{subsect:bar} and \S~\ref{subsect:bar+warp}).}

We have collated the available information of the line of sight velocities from different components of the ISM in order to estimate a generic rotation curve. Although the gas motion may be non-circular and/or not co-planar, a rough estimate of the rotation curve suffices for our qualitative modelling. The line of sight velocity data and the derived rotation curve are shown in Figure~\ref{fig:rotation-curve}.  Data for the line of sight velocities (indicated by cross signs) include $^{12}$CO(2--1) from the circumnuclear gas up to a radius of 0.2~kpc   (\S~\ref{subsect:circumnucleardisk}), $^{12}$CO(2--1) of the gas at larger radii (\S~\ref{subsect:dustlane}), H$\alpha$ \citep{1992ApJ...387..503N} and $^{12}$CO(3--2) \citep{2001AA...371..865L} (Table~\ref{tab-alpha}).  The  estimated rotation curve grows monotonically as 1.2~\kms~pc$^{-1}$ up to r = 0.2~kpc and flattens to 0.2 -- 0.3~\kms~pc$^{-1}$ beyond 1~kpc.  


\subsection{Warped Disk Model}
\label{subsect:warp}
{\color{red}


}

{\bbf We use recent estimates of the inclination and P.A.\ for the
  different tilted rings with radii $r$ $\ge 800$~pc (50\arcsec) from
  \citet{2006ApJ...645.1092Q}, and extend the model for inner radii
  $r$ $<$ 200~pc (12\arcsec) (circumnuclear molecular gas), where we
  use a P.A.\ $\sim 155\arcdeg$ and an inclination $i$ $\sim
  70\arcdeg$ as obtained in \S~\ref{subsect:circumnucleardisk}.
  Originally the model by \citet{2006ApJ...645.1092Q} assumes the
  existence of a circumnuclear disk within $r$ = 100~pc (6\arcsec)
  \citep{2001ApJ...549..915M,1991A&A...245L..13I}, and a gap beyond
  this circumnuclear disk out to a radius $r$ $\simeq$ 800~pc
  (50\arcsec), which is suggested by a comparison with their
  Spitzer/IRAC mid-IR images. Without this gap, the model would show a
  bright linear feature at $r$ $<$ 1~kpc (1 \arcmin) within the
  parallel filaments \citep[][their Fig.~10]{2006ApJ...645.1092Q} that
  is not observed. The infrared surface distribution depends on the
  inclination of the disk, but other mapped species such as the
  Pa$\alpha$ \citep{2000ApJ...528..276M} and our $^{12}$CO(2--1)
  emission show the same trend.  Note that the ad-hoc physical gap
  used in this model implies that there is no physical connection
  between the circumnuclear gas and the warped disk at larger
  radii. 
}

In Figure~\ref{fig:quillen} we show: a) the inclination and P.A.\ as a function of radius that properly reproduce the mid-IR emission, where it is assumed that there is a physical gap in the distribution within 200~pc~$<$~$r$~$<$~800~pc (6\arcsec$<$~$r$~$<$~50\arcsec); b) the projected circular orbits; c) the normalized intensity plot, as obtained from the projected orbit distribution; d) the velocity field map, taking into account the rotation curve in Figure~\ref{fig:rotation-curve}; e) and f) the distribution and velocity field in the inner 1~kpc (1\arcmin).

From a comparison with our  $^{12}$CO(2--1) distribution and velocity field maps (Figure~\ref{fig2}, \ref{fig3}), the overall distribution (parallelogram structure and circumnuclear gas) and kinematics seem to match well, with the circumnuclear gas showing a larger velocity gradient, and the outer gas with a smaller gradual change in velocity. However, a number of details found in our $^{12}$CO(2-1) maps and indicated in \S~\ref{subsect:noncirc} are not well reproduced by this model, including: {\bbf 1) a physical connection between the circumnuclear gas and that at larger radii, 2) brighter SE and NW sides on the parallelogram-shaped 
feature, and 3) an outer curvature of its long sides. 

}

\subsection{Co-planar Weak Bar Model}
\label{subsect:bar}

We inspect whether a disk perturbed by a weak bar potential can
resemble the observed $^{12}$CO(2--1) emission properties in the inner
$\sim$1\arcmin\ (1~kpc). We use a damped orbit model as in
\citet{1994PASJ...46..165W} to describe gas motions in a weak bar
potential within an axisymmetric potential and assuming epicyclic
orbits. {\bbf Note that this model is different to the strong stellar
  bar model (few kiloparsec scale) proposed for Cen~A as suggested by
  \citet{1999A&A...341..667M}, where the observed features in Cen~A
  are even compared to the prototypical barred galaxy
  NGC~1530. Because there is no evidence for a strong stellar
  potential in Cen~A, the assumption that it has a strong bar like
  NGC~1530 may not be well justified. It is also different to the
  models used by \citet{1992ApJ...391..121Q}, where non-circular
  motions within a coplanar disk are obtained as a result of a large
  triaxial stellar potential (i.e. a prolate/triaxial model with axis
  ratios 1:2:2.6). These do not reproduce their $^{12}$CO observations
  at large scale (resolution of about 30\arcsec\ FWHM). We use instead
  a weak non-axisymmetric potential to model the gas only in the inner
  $\sim$1\arcmin\ (1~kpc), which has not been inspected to date for
  this source. }
    
A schematic view of the (projected) features produced by the proposed
weak bi-symmetric potential is shown in Figure~\ref{barmodelscheme},
where all the components (circumnuclear gas -$x_2$ orbits-, as well as
outer gas -$x_1$ orbits- that corresponds to the parallelogram
structure) in the inner 3 kpc of the galaxy are shown.  The
fixed parameters of the weak bar model are: a rotation curve (to fit a
given axisymmetric potential), a bar potential, inclination and
P.A. for the gas disk, bar strength $\epsilon$, bar pattern speed
$\Omega_b$ and a damping parameter $\Lambda$.  {\bbf We use a
  conservative bar strength ($\epsilon$ = 0.03) since there is no
  signature of a strong large stellar bar
  \citep{2006ApJ...645.1092Q}. A larger $\epsilon$ would mainly
  produce a higher contrast between the circumnuclear disk and the
  outer regions.  We estimate the pattern speed $\Omega_b = 0.3$ by
  assuming that the Inner Inner Lindblad Resonance (IILR) is located
  where the circumncuclear gas runs out, at a radius of about $r_{\rm
    IILR} = 0.2$~kpc.  Then the Outer Inner Lindblad Resonance (OILR)
  would be at $r_{\rm OILR} = 0.7$~kpc, approximately at the location
  of the gas at larger radii.  The damping parameter was fixed
  arbitrarily to $\Lambda$ = 0.1, although that only affects the
  orientation of the $x_1$ and $x_2$ orbits
  \citep{1994PASJ...46..165W}. } We use the rotation curve data points
in Figure~\ref{fig:rotation-curve} to obtain a fit (dashed line) using
an axisymmetric logarithmic potential $\Phi_0(r) = 0.5 \times \log(a +
r^2/b)$, with $a = 1.0$ and $b = 0.05$.  We also show in
Figure~\ref{fig:rotation-curve} the curves for $\Omega$ (solid line),
$\kappa$ (dotted line) and $\Omega \pm \frac{\kappa}{2}$
(dashed-dotted lines) as a function of radius.  We inspect a weak bar
model within a flat disk. In Figure~\ref{fig:barPlusquillen2} we show
the following composition: Figure~\ref{fig:barPlusquillen2} a) the
inclination and P.A., which in this case are kept constant at $i$ =
$45\arcdeg$ and P.A. = $100\arcdeg$; Figure~\ref{fig:barPlusquillen2}
b) the damped orbits; Figure~\ref{fig:barPlusquillen2} c) and d) show
the normalized intensity distribution and its corresponding velocity
field; and finally a zoom of these maps to the inner kiloparsec in
Figure~\ref{fig:barPlusquillen2} e) and f). Note that the inclination
is set to a smaller value than its average $i$ $\sim$ 70 --
80\arcdeg\ in order to illustrate the morphology of this model more
clearly.

{\bbf This model produces a disk-like feature with two arms, a large
  concentration at inner radii and a gap outside of these two
  structures, where the number of orbits is at a minimum. Although
  the NW and SE sections of the parallel filaments are brighter than
  the other two sides, the outer curvature and the connections between
  circumnuclear and outer gas could resemble the gas distribution and
  kinematics derived from our maps (\S~\ref{subsect:noncirc}), the
  assumption of a co-planar morphology is not sufficient to reproduce
  the observed parallelogram structure.  

}

\subsection{Combined Model: Weak Bar within a Warped Disk}
\label{subsect:bar+warp}
{\color{red}









%

%
}
{\bbf Constraints to the inclination and position angle (from previous knowledge of the disk properties at larger scale) can be used in order to get a more accurate model based on a weak bar. 
The inclination with respect to the viewer must cross 90\arcdeg\ (edge-on), from $i$ = $70\arcdeg$ inside $r <$ 200 pc (12\arcsec) to $i$ = 70 -- 80\arcdeg\ (but inverted orientation) at $r$ $>$ 1~kpc (1\arcmin) \citep[e.g.][and \S~\ref{jet}]{2006ApJ...645.1092Q}. In the present model the edge-on transition should take place where the lack of emission is more prominent in the pure bar model (\S~\ref{subsect:bar}), at $r$ $\sim$ 600 pc or 40\arcsec, versus the 1 kpc or 60\arcsec\ of the warped disk model (\S~\ref{subsect:warp}).}   
 {\bbf In addition, from our observations, the P.A.\ = 155\arcdeg\ as observed in the circumnuclear gas (r $<$ 200~pc, or 12\arcsec) and the P.A.\ = 120\arcdeg\ in the molecular gas at larger radii ($r > 800$~pc, or $>$ 50\arcsec). 
 We adopt these constraints to the weak bar model in \S~\ref{subsect:bar}}.  
Following the design of Figures~\ref{fig:quillen} (\S~\ref{subsect:warp}) and \ref{fig:barPlusquillen2} (\S~\ref{subsect:bar}), this combined model containing a simple warped model modified by a weak bar model is shown in Figure~\ref{fig:barPlusquillen3}.  Inclination and P.A.\ are detailed in Figure~\ref{fig:barPlusquillen3} a).  The resulting projected orbits are plotted in Figure~\ref{fig:barPlusquillen3} b).  The normalized intensity distribution and the velocity field images, as well as a zoom of these maps are plotted in Figure~\ref{fig:barPlusquillen3} c) -- f). 

{\bbf
 This combined model shows some difference from the
 warped disk model due to the addition of the weak bar.  We find two arms, a large concentration at inner radii and
 a region with a lack of emission outside of these two structures, as in
 \S~\ref{subsect:bar}.  The extra information of the weak bar model
 helps us to naturally explain the connection of circumnuclear and
 outer gas without contradicting the lack of emission along a
 preferential direction, as observed in
 Figure~\ref{fig:barPlusquillen3} b), and in the normalized
 distributions in Figure~\ref{fig:barPlusquillen3} c) and e). In
 addition, the model's spiral arms follow the locations of relatively
 strong $^{12}$CO(2--1) emission to the SE and NW of the center
 (brighter NW and SE regions).  Also, the curvature of the molecular
 gas associated with the parallel lines (see for example
 Figure~\ref{fig2} and Figure~\ref{fig:barPlusquillen3} c) is
 reproduced in our model. On the other hand, a drawback to this
 model is that the velocity field does not show the large velocity
 gradient found in the intersections between the circumnuclear gas and
 the gas at larger radii. 

At larger radii $r > 1$~kpc (1\arcmin), the inclination and P.A. of the rings should follow that of \citet{2006ApJ...645.1092Q}. Our observations can not provide relevant information at these larger scales (or smaller scale, such as the model proposed by \citealt{2007ApJ...671.1329N} for the very inner 3\arcsec, or 50 pc). 

}

\section{Comparisons with Previous Works}
\label{previousWork}

There is a wealth of information available in the literature for Cen~A.  In the following we discuss our results in the context of previous work, emphasizing the molecular gas in the inner kiloparsec of the galaxy, but also taking into account other relevant information to trace the different phases of the ISM that are present in this region. 

\subsection{Molecular Gas}

The main improvement of our SMA  $^{12}$CO(2--1) data over previous single-dish observations is our higher angular resolution, a factor of 45 in beam area with respect to the highest resolution  $^{12}$CO(2--1) maps previously obtained  ($23\arcsec$ resolution or $\sim380$~pc for CO(2--1); \citealt{1993A&A...270L..13R}).  Therefore, our  $^{12}$CO(2--1) data allow us for the first time to distinguish between the different components (i.e. outer and circumnuclear molecular gas and absorption features) and to constrain models more accurately, especially in the interesting nuclear region.



The presence of the ``circumnuclear disk or ring'' (e.g. \citealt{1990A&A...227..342I,1991A&A...245L..13I}; \citealt{1993A&A...270L..13R}; \citealt{1993MNRAS.260..844H}) can be clearly distinguished in our maps.  
This rapidly rotating compact feature was previously determined to have a radius of typically $\sim 100 - 300$~pc, with rotational velocities of $\sim 175$~\kms. 
The estimated extent of the inner disk given by \citet{1991A&A...245L..13I} from the broader component of the CO lines, $30\arcsec$ (or 500~pc), is larger than what we obtain.  \citet{1993A&A...270L..13R} suggested, using beam deconvolution of single-dish observations, that the broad component was in fact a 100~pc edge-on circumnuclear ring.  The extent was therefore underestimated in this case. \citet{1993A&A...270L..13R} suggested that the major axis of the circumnuclear gas has a P.A.\ $\sim 145\arcdeg$, close to our result P.A. = $155\arcdeg$.  

{\bbf Molecular hydrogen observations by \citet{2001ApJ...549..915M}
  (slits along different P.A. using VLT/ISAAC, H$_2$ (J=1--0) S(1)
  2.122 $\mu$m) show the H$_2$ counterpart of our CO emission in the
  inner $r$ = 24\arcsec ~($\sim$ 400~pc), including the slowly rotating
  contribution of the gas in the dust lane, a circumnuclear disk or
  ring ($\sim$ 6\arcsec, or 100~pc) and the nuclear disk ($\sim$
  2\arcsec, 30~pc). The different slits are remarkably similar to our
  P--V diagrams (especially that along P.A. =
  135.5\arcdeg). \citet{2007ApJ...671.1329N} have imaged the same
  transition (H$_2$ (J=1--0) S(1), 2.122 $\mu$m) using VLT/SINFONI) in
  the inner 3$\arcsec$ ($\sim$50 pc). They show that is characterized
  by a mean inclination angle of 34$\arcdeg$ and a P.A. = 155$\arcdeg$
  assuming a warped disk model to describe its gas kinematics. The
  inclination rises from 34.2$\arcdeg$ at 0\farcs76 to $\sim$
  58.9$\arcdeg$ at 1\farcs9, and would probably continue rising to the
  $i$ $\ge$ 70$\arcdeg$ of our circumnuclear
  gas. \citeauthor{2007ApJ...671.1329N}'s model also suggests an
  average P.A. = 155$\arcdeg$, which is the same as the P.A. of our
  $^{12}$CO(2--1) circumnuclear gas. However, the possibility of
  non-circular motions at these scales has not been excluded. }

{\bbf At a larger scale, pure rotational lines of molecular hydrogen H$_2$ (J=2--0) S(0) ($28.22~\mu$m) emission observed with Spitzer/IRS by \citet{2008MNRAS.384.1469Q} are remarkably similar to the $^{12}$CO emission. Both are asymmetric with respect to the parallelogram structure, at least in the inner $45\arcsec$ (or 750~pc) (see their Figure~6), with the SE and NW sides of the parallelogram being brighter than the opposite two sides, as our $^{12}$CO(2--1) maps. 
No explanation for this asymmetry was given by \citet{2008MNRAS.384.1469Q}. A possible interpretation of this has been given in our weak bar within a warped disk model  (\S~\ref{subsect:bar+warp}). Since the molecular hydrogen transition H$_2$ (J=2--0) S(0)  indicates the presence of gas with T $\sim$ 200~K \citep{2008MNRAS.384.1469Q} along the S-shaped feature, our model suggests that this warm molecular gas would be preferentially located along the spiral arms where shocked regions are present. In this scenario the molecular gas would be heated by the abundant star formation and shocks that may take place in these regions.
}

{\color{red} 


}

{\color{red}

}

\subsection{Ionized gas}

{\bbf

The star formation rate in the disk has been estimated to be about
$\sim$ 1 M$_\sun$ yr$^{-1}$, based on the infrared luminosity
estimated by \citet{1990ApJ...365..522E}.
\citet{2000ApJ...528..276M}'s HST/NICMOS observations in a
50\arcsec\ $\times$ 50\arcsec\ circumnuclear region suggest that the
Pa$\alpha$ is located in a ring-like structure with
higher star formation than the rest ($\sim$ 0.3~M$\odot$~yr$^{-1}$). The NW and SE sides of the parallelogram structure are
brighter than their opposite sides in [\ion{Si}{2}] (34.8~$\mu$m),
[\ion{S}{3}] (33.5$\mu$m) and [\ion{Fe}{2}] (25.988$\mu$m)
\citep{2008MNRAS.384.1469Q}, as previously mentioned for the molecular
gas.  These species likely trace HII regions.  Other higher ionization
tracers such as [\ion{O}{4}] (25.9$\mu$m) and [\ion{Ne}{5}]
(24.3$\mu$m) are distributed along the jet. This would be in agreement
with non-rotational motions exhibited in the inner 3\arcsec\ in
Br$\gamma$, [\ion{Fe}{2}] and [\ion{Si}{4}]
\citep{2007ApJ...671.1329N}, and likely result from interaction of
the gas with the jet.

The reason for the star formation regions not being equally
distributed along the two parallel-shaped structure is still unknown.
Variations in the brightness of star forming regions could be partly
responsible for this effect, but since the molecular gas also shows the same trend, it
may be more appropriate to suggest that these are in fact regions with
larger amounts of gas and with enhanced star formation, probably as a
result of shocks driven into the gas by the putative weak bar.

}

\subsection{Dust Emission}
\label{previousWork-1}

{\color{red}












         }

We compare the distribution of the molecular gas as traced by the
$^{12}$CO(2--1) line to the dust emission.  Resolutions
spanning $3-6\arcsec$ are available in $5 ~\mu$m  -- 18~$\mu$m ISOCAM
images for the warm dust (including complex molecules) by
\citet{1999A&A...341..667M}, covering a 2\arcmin\ field around the
core.  Best resolutions, of the order of $1\arcsec-2\arcsec$ over a
5\arcmin $\times$ 5\arcmin\ region, have been recently obtained by
\citet{2006ApJ...645.1092Q} using Spitzer/IRAC in the $2~\mu$m --
8~$\mu$m range.  For the colder dust, $450~\mu$m and $850~\mu$m
submillimeter observations have been obtained using SCUBA with HPBWs
of $14\farcs5$ and $8\farcs0$ covering the central 450\arcsec $\times$
100\arcsec , respectively \citep{2002ApJ...565..131L}.  While the
$5~\mu$m -- 18~$\mu$m ISOCAM images and submm images show an S-shaped
feature and an extension with P.A. = 145$\arcdeg$ in the inner $r$ =
15\arcsec , the Spitzer/IRAC $2~\mu$m -- 8~$\mu$m images show a
structure that looks like a parallelogram structure, without a
clear extension to the nuclear region.

The warm dust and emission by complex molecules (PAHs) are well
coupled to the CO(2--1) distribution, both show a similar structure,
and even the main clumps are collocated.  In Figure~\ref{fig:mom0-mir}
we show an overlay of the mid-IR $8~\mu$m image from Spitzer
\citep{2006ApJ...645.1092Q} and our $^{12}$CO(2--1) line distribution.
The agreement between the mid-IR and CO line emission is excellent,
except for the very center of the galaxy.  The dust emission (as
traced by mid-IR) seems to be depleted at intermediate
radii: \citet{2006ApJ...645.1092Q} proposed a gap from $6\arcsec$~$<$~$r$~$<$~
$50\arcsec$ (100~pc~$<$~$r$~$<$~800~pc) in order to match their optically
thin warped disk model to the mid-IR emission. {\bbf However, since the
  infrared surface brightness depends on the inclination of the disk,
  \citet{2006ApJ...645.1092Q} noted that if the disk twists to lower
  inclinations at small radii it would have lower surface brightness
  near the nucleus.}  \citet{1992ApJ...387..503N} also suggest that
there is a lack of emission in ionized gas within $60\arcsec$ {\bbf
  although this could be partly caused by extinction.  Assuming a
  warped disk model, our observations would confirm the existence of a gap
  between 200 $<$ $r$ $<$ 800~pc and a central
  concentration at $r < 200$~pc (12\farcs5). However, it would not explain the
  obvious physical connection between the circumnuclear gas and gas at
  larger radii, which would be more easily explained by the combined
  model (\S~\ref{subsect:bar+warp}).}

{\color{red}
}

We show in Figure~\ref{fig5} the ratio of emission at mid-IR $8~\mu$m
\citep{2006ApJ...645.1092Q} and $^{12}$CO(2--1) emission.  It is observed that toward the center the ratio of
$^{12}$CO(2--1) emission to $8~\mu$m is larger than in
the outer regions.  The observed gradient might be explained by
temperature differences toward the center caused by winds and
high-energy radiation from the AGN (\citealt{1998A&ARv...8..237I} and
references therein), as derived by: i) the very red near-IR colours
within the inner $r$ = 200 pc (12\arcsec), which cannot be explained
just by extinction, but also need elevated dust temperatures, shocks
and UV and X-ray radiation from the nucleus; and ii) the compactness
of [\ion{Fe}{2}], Pa$\alpha$ and H$_2$ emission, with extents of $\sim
40$~pc and rotating with 215~\kms\ at temperatures of $\sim$1000~K and
densities in excess of $10^5$ cm$^{-3}$.

\subsection{Geometrical Relation between Circumnuclear Gas and Nuclear Activity}
\label{jet}

{\color{red}







}

A progression from the disk-like structure along the dust lane (few kpc scale) to the circumnuclear gas (few hundred pc scale) and then to the inner accretion disk (inner 10~pc) would be expected if this gas is indeed the fuel for the AGN.  In this case the accretion disk might be expected to be oriented perpendicular to the jet.  However, this has been the object of controversy over the last two decades.  While in spiral galaxies inner disks have a random orientation with respect to the large scale outer disks \citep{1984ApJ...285..439U}, this is probably not the case in elliptical galaxies \citep{1992A&A...255...35M}.  However \citet{2002ApJ...575..150S} find that this correlation is not strong using three dimensional information.  Recently \citet{2005A&A...435...43V} have also shown, using three dimensional information, that the jet is approximately perpendicular to the dust structure in the galaxies possessing filamentary dust structures or dust lanes (only $\simeq20\arcdeg$ misalignment angle, as is the case of Cen~A), whereas galaxies having regular dust structures with an elliptical appearance have a much wider distribution ($\simeq45\arcdeg$ misalignment angle).

To illustrate that the circumnuclear gas is indeed perpendicular to the inner jet as projected in the sky plane we show in Figure~\ref{fig:mom0-mir} a superposition of the 21~cm radio continuum image \citep{2002ApJ...564..696S} and our  $^{12}$CO(2--1) map.  The relativistic jet is at a P.A.\ $\simeq 51\arcdeg$, which is nearly perpendicular in projection to the circumnuclear gas, P.A.\ $\simeq 155\arcdeg$.  At an even smaller scale, the radio continuum emission at 2.3, 4.8, 8.4 and 22.2~GHz observed using the Southern Hemisphere VLBI and VLBA by \citet{1998AJ....115..960T}, or even the space-VLBI at 4.9 GHz by \citet{2006PASJ...58..211H}, nicely show the same P.A. $\simeq$  51\arcdeg\ for the nuclear jet.  An X-ray jet is also reported to be coextensive at a similar P.A.\ \citep{2002ApJ...569...54K}.

Information about the three dimensional geometry of the jet can be extracted from the difference in luminosity of the NE jet and the SW counter jet caused by the relativistic beaming effect and the motions and temporal variations in the jets.  From Figure~\ref{fig:mom0-mir} b), it is clear that the NE inner jet is brighter than the SW counter jet and thus the former is likely pointing towards us. \citet{1998AJ....115..960T} conclude that the axis of the Cen~A jet lies between $i$ $\simeq 50\arcdeg$ and $80\arcdeg$ to our line of sight.  
We derived an inclination for the circumnuclear gas of  $i$ $\ge 70\arcdeg$ (\S~\ref{result}) if the near side is to the SW.  If the nuclear jet happens to be perpendicular to the circumnuclear gas, an orientation of the nuclear jet with respect to the line of sight of $\simeq 50\arcdeg - 80\arcdeg$ \citep[][using VLBI observations]{1998AJ....115..960T} would be compatible.
{\bbf This is in contrast to the results of \citet{2007ApJ...671.1329N}, whose mean inclination of the modelled warped disk is $i$ = 34$\arcdeg$.
This value is closer to the 20\arcdeg\ $<$ $i$ $<$ 50\arcdeg\ found by \citet{2003ApJ...593..169H} using VLA observations. The VLBI observations by \citeauthor{1998AJ....115..960T} have much higher angular resolution and trace only the innermost region of the jet.

}

 We can also estimate the time scale in which the observed amount of
 circumnuclear molecular gas will be consumed just by accretion with
 the current rate of energy emitted by the nucleus. The active nucleus
 is well below the Eddington luminosity
 \citep{2001ApJ...549..915M}. The bolometric luminosity is estimated
 to be $L_{bol}$ $\simeq$ 10$^{43}$ erg s$^{-1}$, with half of it
 taking place at high energies \citep{1998A&ARv...8..237I}. With a
 typical efficiency from the total luminosity to mass accretion of
 0.1, we obtain a mass accretion rate of dM/dt $\sim$ 2 $\times$
 10$^{-3}$ M$_\sun$ yr$^{-1}$.  With our derived total amount of
 molecular gas in the nuclear region M$^c_{gas}$ = (0.8 $\pm$ 0.1)
 $\times$ 10$^7$ M$_\sun$ (with a conversion factor $X$ = 0.4
 cm$^{-2}$ (K km s$^{-1}$)$^{-1}$), the time scale to consume it would
 be of the order of 1 Gyr.  Note however that the Eddington luminosity
 estimated with a black hole mass of M$_{BH}$ $\simeq$ 10$^{7-8}$
 M$_\sun$
 \citep{2001ApJ...549..915M,2006A&A...448..921M,2007MNRAS.374..385K,2007ApJ...671.1329N}
 is about 2--3 orders of magnitude larger than the measured
 $L_{bol}$. The entirety of the circumnuclear molecular gas would be consumed
 if the Eddington luminosity is maintained within a range of
 10$^{6-7}$ yr.
  

\section{Discussion}
\label{discussion}


\subsection{Warped Disk or Combined model?}
{\color{red}


       }
{\bbf The warped disk model is the preferred model to date. It has
  been used successfully to reproduce the observations both at a large scale \citep[e.g.][]{2006ApJ...645.1092Q} and also at a
  few parsec scale \citep{2007ApJ...671.1329N}. Our observations give
  information about the missing intermediate scales (100 pc to 1 kpc).
  The warped disk model suggests that there is a gap in the
  distribution at $6\arcsec$~$<$~$r$~$<$~50\arcsec , which is introduced
  ad-hoc in order to reproduce the observed lack of mid-IR emission between the long sides of the parallelogram
  filaments \citep{2006ApJ...645.1092Q}, although it is not easy
  to distinguish whether this is just an inclination effect on the surface
  brightness of the dust emission.
  Our $^{12}$CO(2--1) data, which is not affected by such effect, 
  firmly show that there is still a lack of emission. If this model is correct, this would confirm
  the existence of the physical gap.
  
  However, the warped disk model does not explain a number of deviations
  from our $^{12}$CO(2--1) maps (\S~\ref{subsect:noncirc}), such as the brighter
  sides on the NW and SE of the parallel filaments. Furthermore, the P--V diagram along P.A.\ $\sim135\arcdeg$ provides
  seemingly clear evidence that there are continuous layers of molecular gas
  from the innermost region in the SE to the outer disk (southern
  parallel filament), which is not in agreement with a gap within  6\arcsec ~$<$~$r$~$<$~50\arcsec ~ (100~pc~$<$~$r$~$<$~800~pc) in the CO
  distribution as assumed in the warped disk model.} 
 Twin peaks in CO emission also
seem to be observed into the nuclear region of Cen~A
(\S~\ref{subsect:noncirc}), which are usually detected in barred
spiral galaxies
(e.g. \citealt{1992ApJ...395L..79K,1996ApJ...461..186D}).  Molecular
gas, ionized gas and dust seem to be coextensive
(\S~\ref{previousWork}). Most of the species show brighter emission to
the NW and SE, rather than spread over the parallelogram-shaped
structure.  
 Overall, in a qualitative way, the presented contribution of a weak bar potential  (see \S~\ref{subsect:bar}, \ref{subsect:bar+warp}) reproduces the brighter emission of the NW and SE sides (\S~\ref{previousWork}) with enhanced SF, and naturally explains the formation of circumnuclear gas, its connection to gas at larger radii and a lack of emission along a preferential direction between the two. This weak bar could have originated in the merger event that brought the gas into the dust lane,  and would efficiently help to
   funnel gas into the circumnuclear environment.


Note that this model is used just to illustrate the possible contribution of a weak bar model in Cen~A. 
First, the pattern speed has been estimated by the unknown location of the IILR, which was set to the edge of the circumnuclear gas.  
Also, the model does not specify the distribution of orbit density (gas surface density) in the radial direction, and the gas distribution in the tangential direction is bisymmetric by definition even though many barred galaxies show deviations from it. And finally, the model is applicable only for infinitesimally weak bars.
High resolution observations of CO transition lines along the whole dust lane will help to determine which model (or combination) is more likely. 
If confirmed that a weak bi-symmetric potential is present in the central region of Cen~A, one can wonder whether they could be one of the main drivers of activity, even in elliptical galaxies like Cen~A. Otherwise one would have to argue that viscosity within the disk may be important to transport the gas toward the nuclear region \citep{2000A&A...357.1123D}. 

\subsection{Relation between Molecular Gas Properties and Nuclear Activity}

The gas-to-dynamical mass ratio and the degree of gas concentration toward the center are important parameters to know the star formation activity and the possible gas infall rate along a bar to a galactic center \citep[e.g.][]{1999ApJS..124..403S}.
We compare these two properties for Cen~A with the sample of spiral galaxies in \citet{1999ApJS..124..403S}.  \citet{1999ApJS..124..403S} set a limit of r = 500 pc between circumnuclear gas and outer gas. In order to compare with \citet{1999ApJS..124..403S}, we use use a constant X = 3.0 $\times$ 10$^{20}$~cm$^{-2}$~(K \kms)$^{-1}$ for both external and nuclear molecular gas.
Although these two properties are affected by the uncertain X-factors as a function of radius within a galaxy, we assume that this source of error is similar for different galaxies, and thus the comparison still allows us to obtain meaningful results.

$M_{\rm gas}$ (gas mass) in the inner 500 pc and the surface density $\Sigma_{\rm gas}$ will also depend on how the molecular gas is distributed, which is different for each of the models explained in \S~\ref{model}.  In case of the warped disk model, the gas mass and the surface density within 500~pc are $M_{\rm gas, 500~pc} \sim 6.1 \times 10^{7}$~M$_{\odot}$ and $\Sigma_{\rm gas, 500~pc} \sim 78$~M$_\odot$~pc$^{-2}$.  In case of the weak bar models, most of the molecular gas detected with our observation is within the radius of 500~pc, so $M_{\rm gas, 500~pc} \sim 1.8 \times 10^{8}$~M$_{\odot}$ and $\Sigma_{\rm gas, 500~pc} \sim 230$~M$_\odot$~pc$^{-2}$.


The dynamical mass in the central 500~pc radius is calculated as
$M_{\rm dyn, 500~pc}$ = $r (v / \sin i)^2 / G$ = $4.6 \times
10^9$~M$_{\odot}$, obtained from the maximum rotational velocity $v
\simeq 200$~km~s$^{-1}$ and inclination of $i$ = $70\arcdeg$.  The
gas-to-dynamical mass ratio, $M_{\rm gas, 500~pc} / M_{\rm dyn,
  500~pc}$, for the warped disk and the weak bar models can be
calculated as 0.01 and 0.04, respectively.  Both of these values show
that the gas mass at the center of Cen~A is only a few percent of the
dynamical mass (See \S~\ref{sub:moleculargasmass}).  If we compare the
results of \citet{1999ApJS..124..403S}, Cen~A appears around the low
end of their sample in either model.  The galaxies with low
gas-to-dynamical mass ratio in the \citeauthor{1999ApJS..124..403S}
sample are dominated by Seyfert galaxies, and this is consistent with
the existence of an AGN at the center of Cen~A (although AGN activity
is much higher in Cen~A than in Seyferts).  The low gas-to-dynamical
mass ratio also suggests a low rate of star formation
\citep{1999ApJS..124..403S}, but it is inconsistent with the results
by \citet{2000A&A...359..483W}, who find that the dense gas fraction
$L_{\rm HCN} / L_{\rm CO}$ as well as the star formation efficiency
$L_{\rm FIR} / L_{\rm CO}$ in the nuclear region of Cen~A is
comparable to ultra-luminous infrared galaxies (ULIRGs).  This
inconsistency may be, however, due to the effect of the AGN, not due
to the star formation activities.  Recent interferometric studies of
Seyfert galaxies show that the HCN line is strongly enhanced toward
the AGN dominated nuclei
\citep[e.g.][]{2001ASPC..249..672K,2005AIPC..783..203K}.  Dust around
the nucleus can also be heated by the AGN, and it can enhance $L_{\rm
  FIR}$.  Future higher spatial resolution CO and HCN observations,
and comparisons with high resolution infrared data, will address this discrepancy.


To calculate the degree of gas concentration, we first calculate the gas surface density within the isophotal radius at 25~mag arcsec$^{-2}$, $R_{25}$, in B-band, following \citet{1999ApJS..124..403S}.  Since this is an elliptical galaxy and the molecular gas is elongated along the minor axis, we take the $R_{25}$ along the minor axis.  In this case, $R_{25} = 10.0\arcmin = 10.0$~kpc.  This yields $\Sigma_{\rm gas, R_{25}} \sim 1.5$~M$_\odot$~pc$^{-2}$.  The degree of gas concentration is therefore calculated as $f_{\rm con} = \Sigma_{\rm gas, 500~pc} / \Sigma_{\rm gas, R_{25}} \sim 52$ for the warped model and 150 for the weak bar models ($\sim$10 and $\sim$30 respectively if a ratio of external to internal X factor of 5 is taken into account, see \S~\ref{sub:moleculargasmass}). The value for the warped disk model is located at the high end of non-barred galaxies in the \citeauthor{1999ApJS..124..403S} sample, and that with the contribution of a weak bar model is located at the low end of barred galaxies.  This result shows that the degree of gas concentration of Cen~A in each of the models is consistent with the properties of disk galaxies which is surprising given the elliptical nature of Cen~A, and probably that the concentration is not as extreme as in large-scale barred galaxies, as expected for a weak stellar potential proposed here.

\section{Summary and conclusions}
\label{conclusion}

{\bbf 
We have observed the molecular gas as traced by the  $^{12}$CO(2--1) line in the central 1~kpc of the nearby elliptical galaxy Cen~A.  The angular resolution achieved using the SMA is  6\farcs0 $\times$ 2\farcs4 (100~pc $\times$ 40~pc). Our main findings are summarized as follows:

\begin{itemize}
\item We can distinguish between the following components: i) a rapidly rotating circumnuclear gas disk or torus  (broad plateau component of the emission line) of 400~pc extent (24\arcsec), with a P.A.\ = $155\arcdeg$ and perpendicular to the jet, and ii) molecular gas at larger radii associated with two apparently parallel filaments at a P.A.\ = $120\arcdeg$ (narrow line component). The latter is observed to be coextensive with other components of the interstellar medium, such as the dust emission and PAHs from the 8$\mu$m mid-IR continuum emission, and is associated to the dust lane.

\item We find the following signatures for non co-planar or non-circular motions: i) a large scale warp  in the two parallel filaments, ii)  brighter emission on the SE and NW sides in the two parallel filaments, iii) outer curvature in these filaments, iv) connections between the circumnuclear gas and that at larger radii, v) asymmetries in the distribution and kinematics in the circumnuclear gas.  A twin peak CO emission line is probably observed in the circumnuclear gas. The P--V diagram shows a very steep rise with $dv/dr = 19.2$~\kms~arcsec$^{-1}$ within the inner 200~pc (12\arcsec),  six times larger than in the outer component. 

\item We extend the currently prevalent warped-thin disk model in order to match the details in the inner kiloparsec of our  $^{12}$CO(2--1) maps that still remained unexplained, in particular the contribution of the circumnuclear gas. However, deviations from this model include  the brighter emission on the SE and NW sides in the two parallel filaments and the physical connection between the circumnuclear gas and that at larger radii.  The contribution of a weak bar model is used to explain these deviations from the warped disk model. In addition to this, it naturally explains the rapidly rotating circumnuclear region, the observed connections between the outer gas and circumnuclear gas, and the gap feature between them. In this scenario the NW and SE sides of the filaments would be the shocks where larger concentrations of molecular gas are located, as well as enhanced SF as has been observed by other authors.

\item We find a low gas-to-dynamical mass ratio, suggesting a low rate of star formation close to the AGN. The  molecular gas concentration index $f_{\rm con} = \Sigma_{\rm gas, 500~pc} / \Sigma_{\rm gas, R_{25}} \sim 52$ for the warped model and 150 for the weak bar model. 
When compared with disk galaxies it is observed that the degree of gas concentration of Cen A (for both models) is consistent with the properties of disk galaxies 
which is surprising given the elliptical nature of Cen~A. 

\end{itemize}

}

\acknowledgments{

We thank the SMA staff members who made the observations reported here possible. We especially acknowledge K. Sakamoto, P. T. P. Ho and L. Verdes-Montenegro for useful comments on the manuscript. We also thank A.\ Quillen for kindly making available the mid-IR images from Spitzer/IRAC and IRS and A. Sarma for the 21cm continuum data from VLA.  This research has made use of NASA's Astrophysics Data System Bibliographic Services, and has also made use of the NASA/IPAC Extragalactic Database (NED) which is operated by the Jet Propulsion Laboratory, California Institute of Technology, under contract with the National Aeronautics and Space Administration. This research was supported by a Marie Curie International Fellowship within the 6$\rm ^{th}$ European Community Framework Programme.
}

{\it Facilities:} \facility{SMA}

\bibliography{cena}

\clearpage


{\color{red}

}

\begin{figure}
\epsscale{0.6}
\plotone{f1.eps}
\caption{ Comparison of the SEST spectrum from \citet[; dashed line]{1992A&A...265..487I}  with that obtained by us using the SMA (solid line) and convolved to the 23\arcsec\ beam of the SEST. Note that the high velocity wings (300 to 500 \kms\ and 600 to 800 \kms) show good agreement, while we miss some flux in the narrow line (500 \kms\ to 600 \kms ). The small deviation in the broad plateau is probably due to a slightly wrong baseline subtraction in the single-dish data. 
\label{fig-compara-spectra}}
\end{figure}

\begin{figure}
\epsscale{1.1}
\plotone{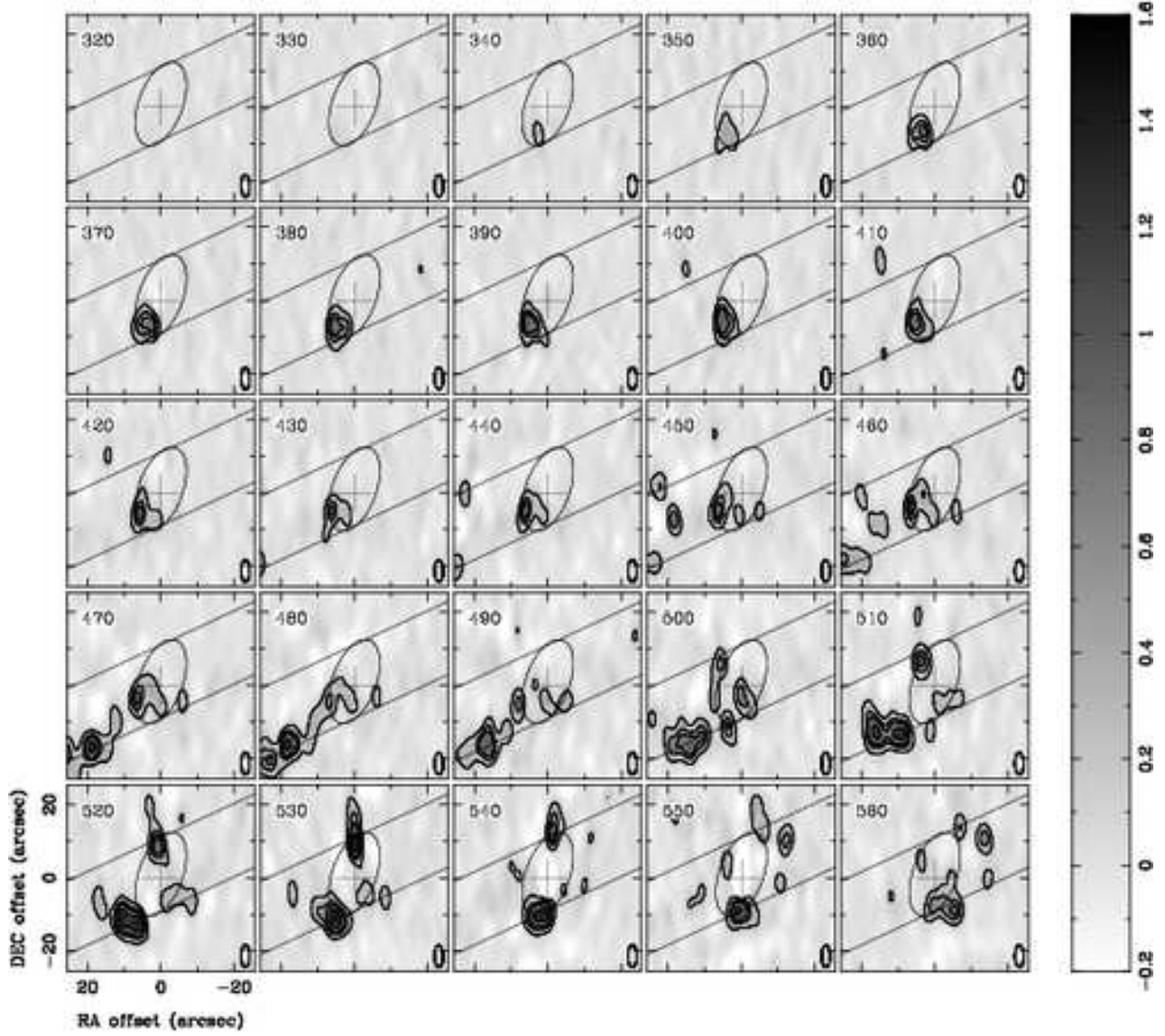}
\caption{Channel map of the  $^{12}$CO(2--1) line of Cen~A, made using uniform weighting, in the LSR velocity range V = $320 - 810$~\kms\ in 10~km~s$^{-1}$ bins.  The velocities are shown in the upper left corner and the synthesized beam is shown in the lower right corner of each panel.  The rms noise of an individual channel is 64~mJy~beam$^{-1}$.  The contour levels are 3, 7, 16 and 25$\sigma$ (which corresponds to 0.19, 0.44, 1.02 and 1.60~Jy~beam$^{-1}$).
The cross sign shows the position of the AGN: R.A.\ = 13${\rm ^h}$25${\rm ^m}$27.${\rm^s}$615 ; Dec. =  --43${\rm ^o}$01$\arcmin$08\farcs805 \citep{1998AJ....116..516M}. 
The central ellipse indicates the location of the resolved inner circumnuclear component, and the two parallel lines the location of the molecular gas associated with the dust lane (narrow line component).  Note that the P.A.\ of the major axis of the ellipse (P.A.\ $\simeq 155\arcdeg$) is different from that of the gas at larger radii (P.A.\ $\simeq 120\arcdeg$). Absorption features are found toward the galactic center, which are apparent in our maps in the $540 - 550$~\kms\ channels but are also present up to about 625~\kms\ \citep{2008ApJ...000..000E}.
\label{fig1}}
\end{figure}

\setcounter{figure}{+1}
\begin{figure}
\epsscale{1.1}
\plotone{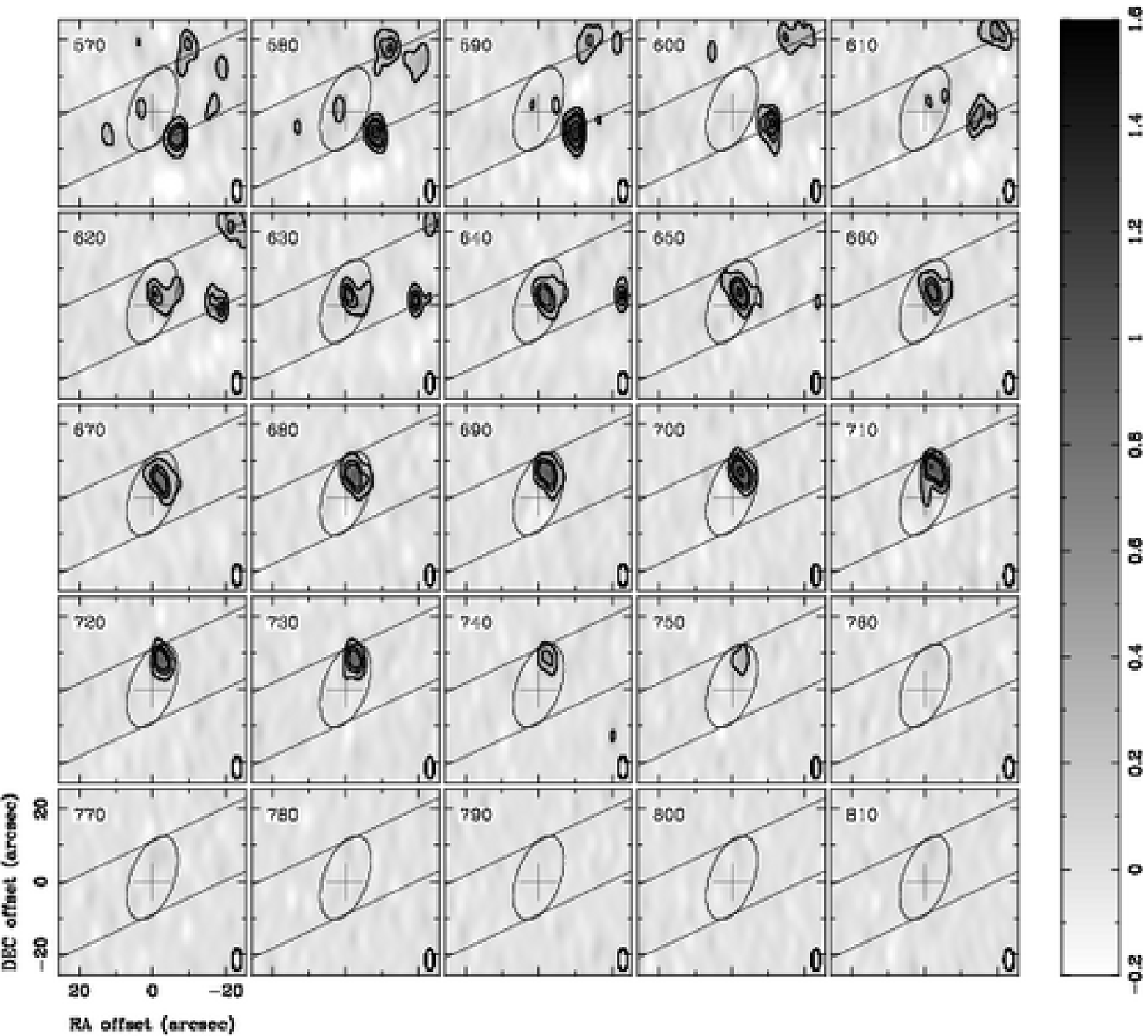}
\caption{Continued.}
\end{figure}

\begin{figure}
\begin{center}
\includegraphics[width=12cm]{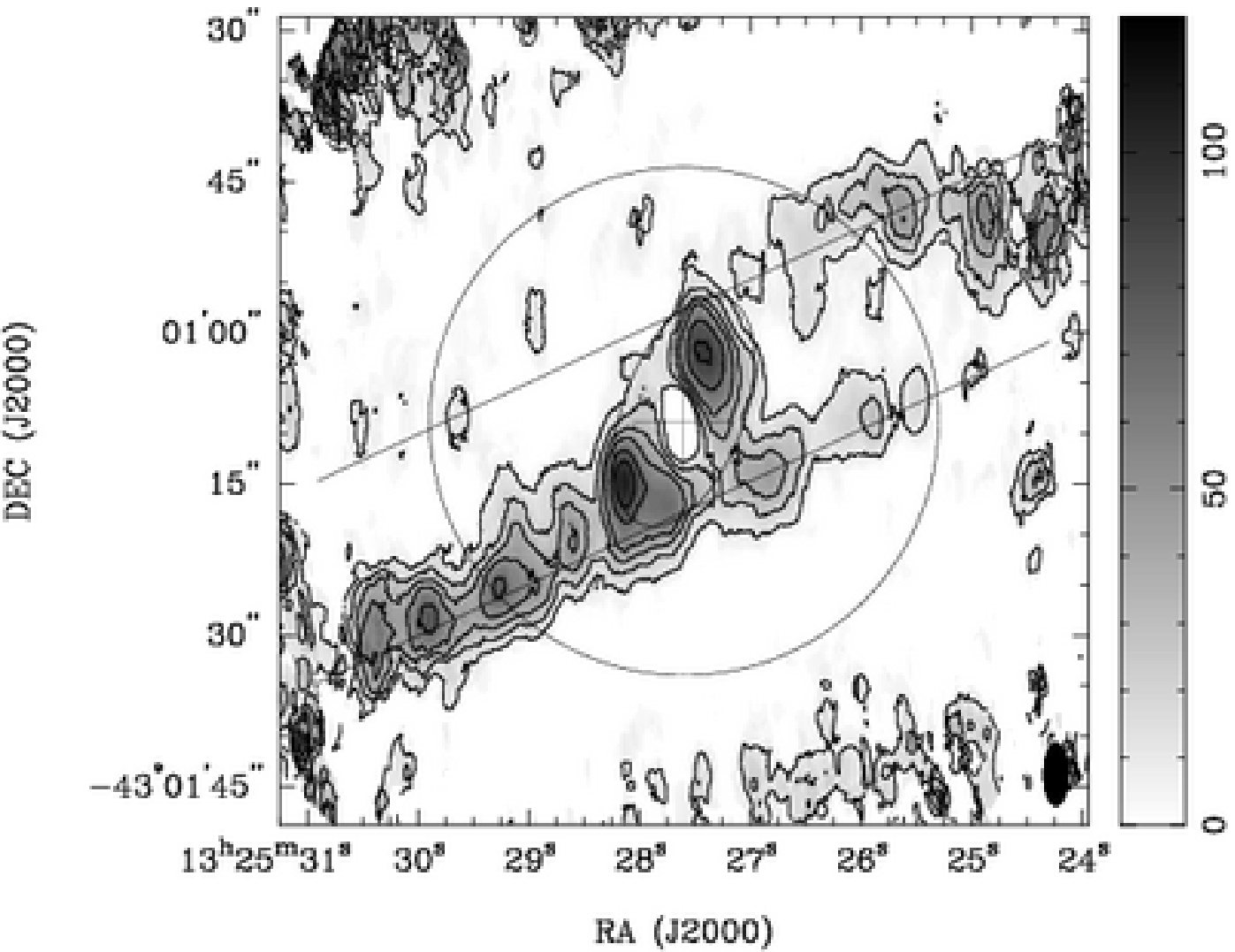}\\
\end{center}
\caption{ $^{12}$CO(2--1) integrated intensity map in Cen~A.  Primary beam correction was performed in this image, as emission was found beyond the $52\arcsec$ full half power width of the primary beam (circle).  Contour levels are 22.5, 36, 49.5, 72, 90 and 112.5~Jy~beam$^{-1}$ km s$^{-1}$.  The synthesized beam ($2\farcs4 \times 6\farcs0$) is indicated by a filled ellipse in the lower corner of the plot, and the colour scale for the  $^{12}$CO(2--1) intensity is shown beside the plot in Jy~beam$^{-1}$ \kms .  For the central ellipse, the cross, and the two lines, see Figure~\ref{fig1}.
\label{fig2}}
\end{figure}

\begin{figure}
\begin{center}
\includegraphics[width=12cm]{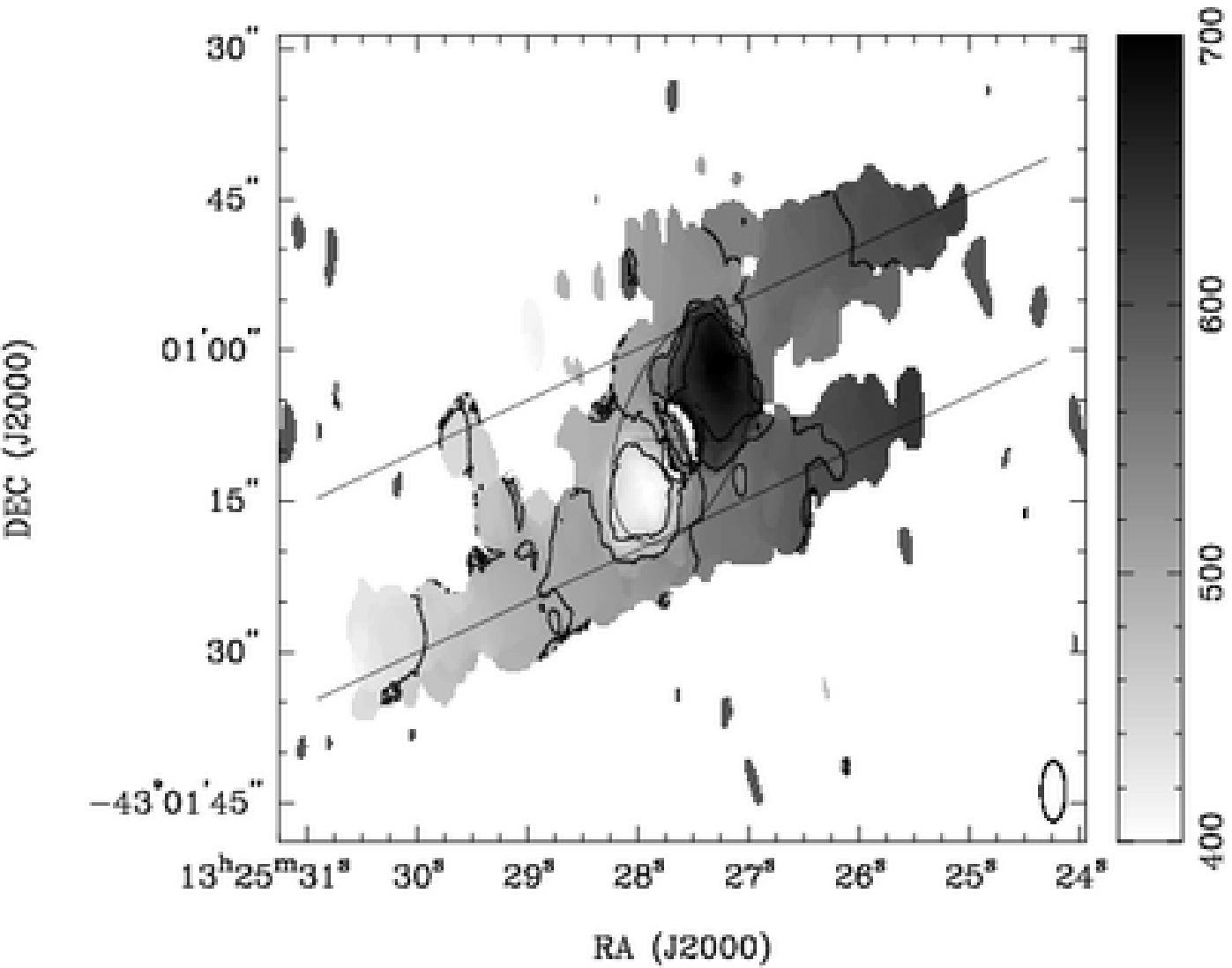}
\end{center}
\caption{ $^{12}$CO(2--1) (intensity weighted) velocity field map of Cen~A.  Contours are placed every 50~km~s$^{-1}$, from $425 - 675$~km~s$^{-1}$.  The color scale ranges from 400 \kms\ up to 700 \kms . The size of the synthesized beam is shown in the lower right corner.  For the central ellipse, the cross, and the two lines, see Figure~\ref{fig1}. 
\label{fig3}}
\end{figure}

\begin{figure}

\includegraphics[width=16cm]{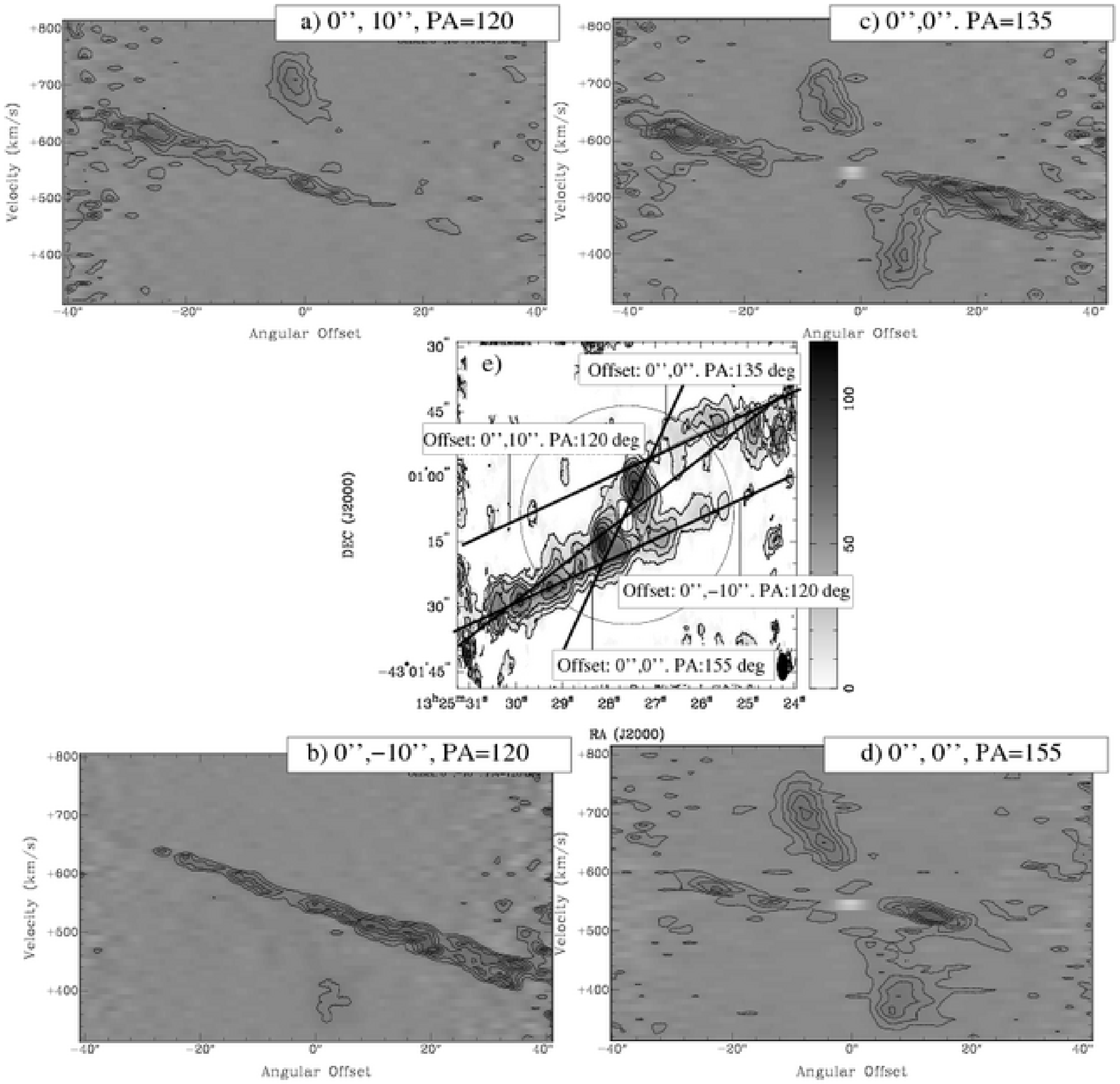}
\begin{center}
\end{center}
\caption{ a) -- b) Position--Velocity (P--V) diagrams at $120\arcdeg$ (dust lane major axis) with an offset of $\pm10\arcsec$ in Dec., with contour levels  0.5, 0.9, 1.3, 1.7 and 2.1~Jy~beam$^{-1}$.) c) -- d) P--V diagrams at P.A.\ = $135\arcdeg$ and  $155\arcdeg$ (circumnuclear gas major axis), both centered in the AGN. Contour levels are: 0.4, 0.7, 0.9, 1.2, 1.5 and 1.8~Jy~beam$^{-1}$. e) Plot showing the different P--V cuts over the  $^{12}$CO(2--1) integrated intensity map.
Positive angular offsets are given for the SE, negative for the NW.
\label{fig-pv}}
\end{figure}

\begin{figure}
\centering
\includegraphics[width=12cm]{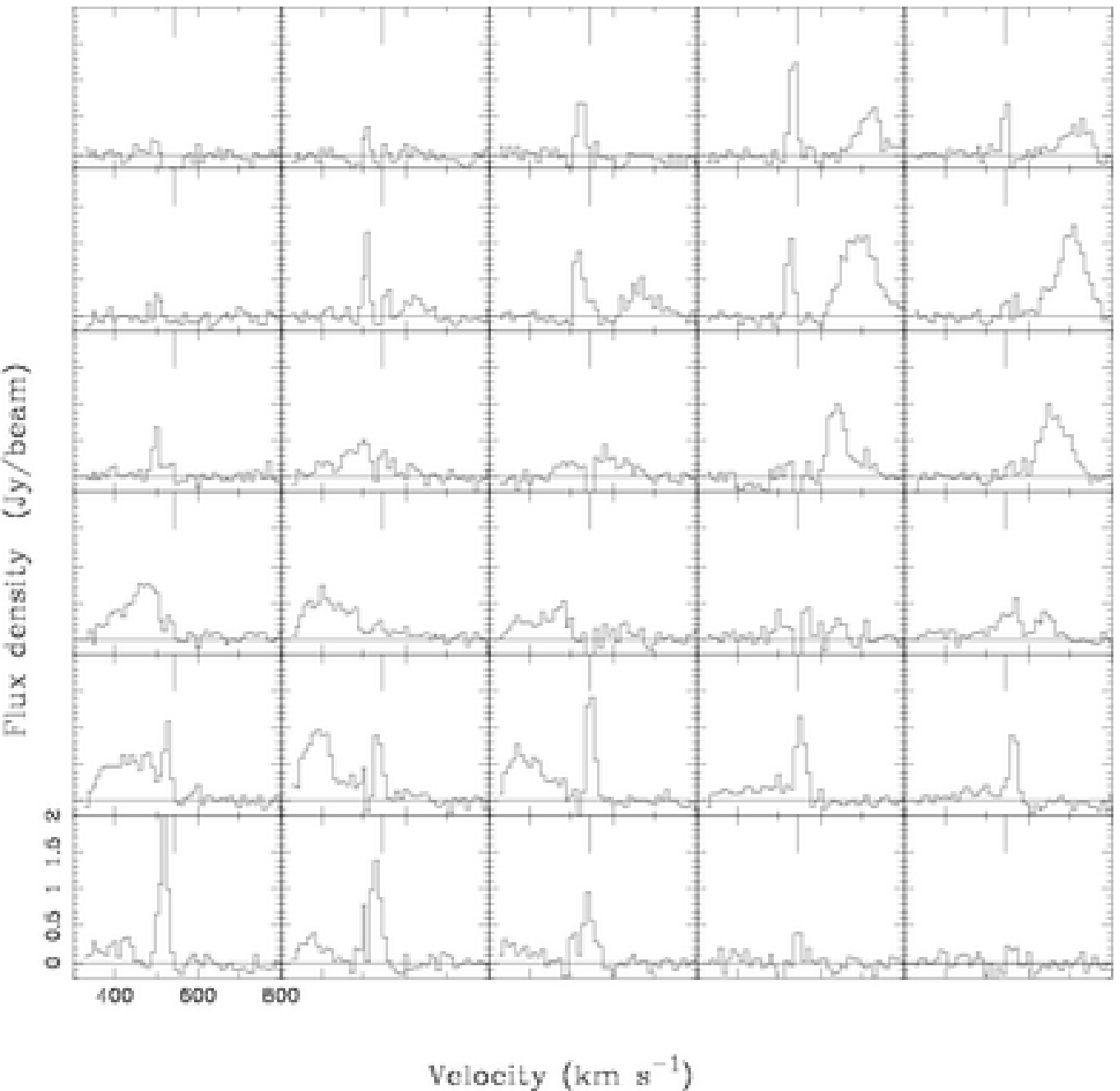}\\
\includegraphics[width=8cm]{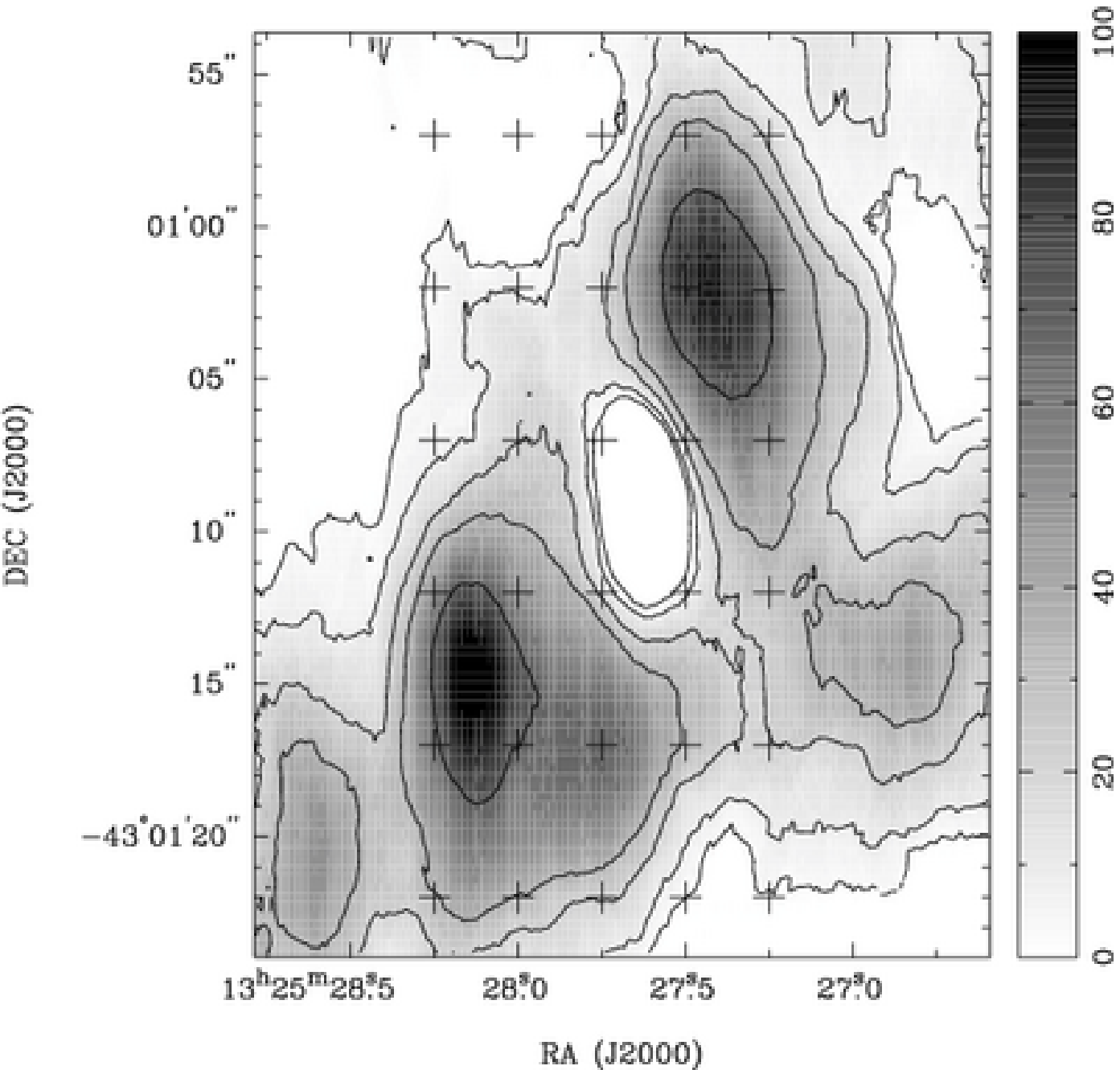}

\caption{a) $^{12}$CO(2--1) line profiles on a  5$\times$6 grid with spacings $0\fs25$ in R.A.\ and $5\arcsec$ in Dec., covering the circumnuclear disk and part of the outer molecular gas. The flux density of the profiles spans from 0 to 2~Jy~beam$^{-1}$, and the x-axis shows velocity from 320 to 810~\kms\ with 10~\kms\ resolution.  b) Integrated  $^{12}$CO(2--1) flux densities map (see Figure~\ref{fig2}). Cross marks the grid positions shown in Figure~\label{fig-spectrumComposition} a).
\label{fig-spectrumComposition}}
\end{figure}

\begin{figure}
\centering
\includegraphics[width=12cm,angle=-90]{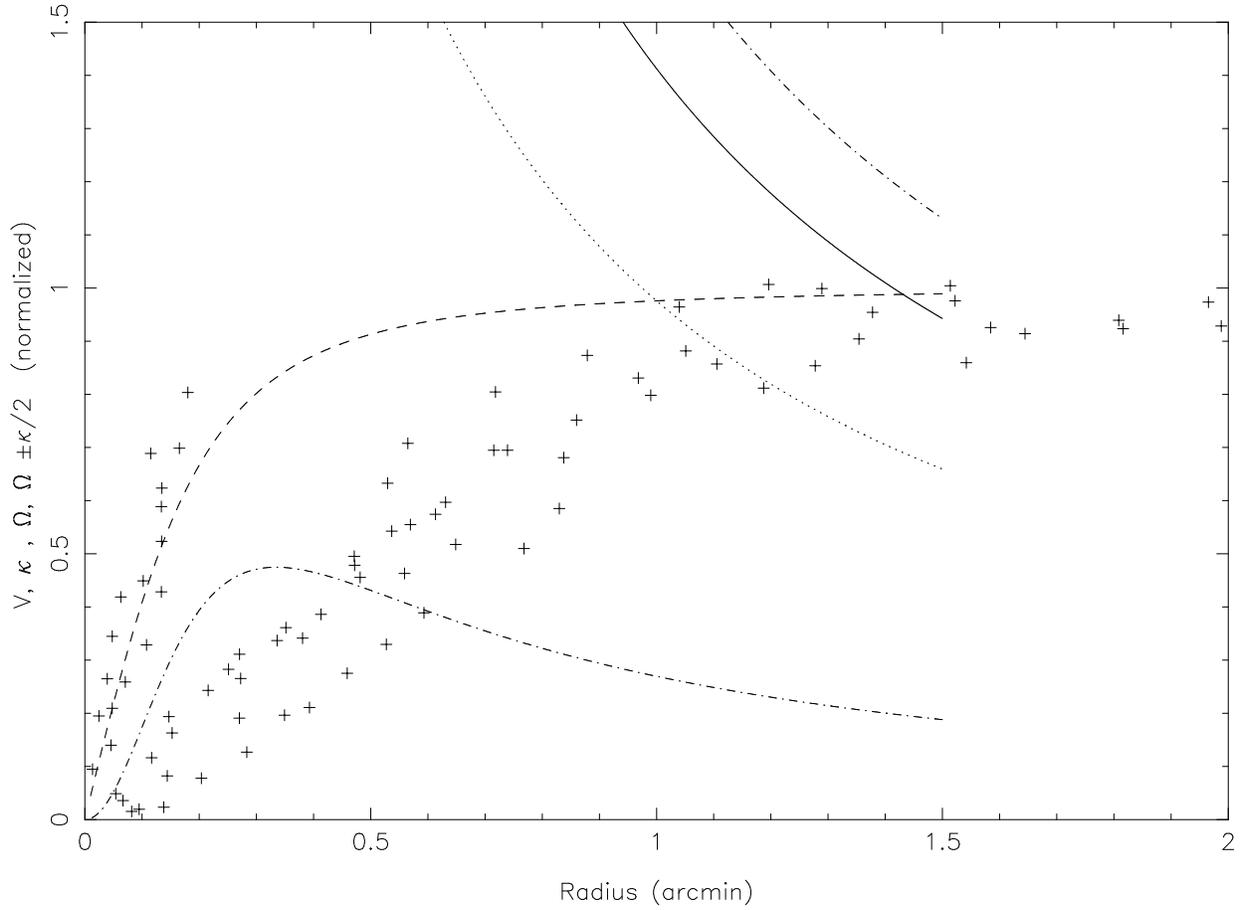}
\caption{Rotation curve (dashed line) roughly estimated using line of sight velocity data (cross signs) of  $^{12}$CO(2--1) in the high velocity circumnuclear gas up to $r$ = 200~pc (12\arcsec , this work),  $^{12}$CO(2--1) in the outer regions (this work), H$\alpha$ (Nicholson et al. 1992) and  $^{12}$CO(3--2) \citep{2001AA...371..865L}. Data points are obtained from the derived line of nodes or directly from the position velocity diagrams, overlapping both the receding and approaching sides, {\bbf divided by $V$ = 260 \kms, which corresponds to its maximum}. The bimodality of the inner data points corresponds to the nuclear and outer gas. A reasonable fit which suffice to qualitatively model the nuclear region of the galaxy is  obtained using an axisymmetric logarithmic potential $\Phi _0(r) = 0.5 \times log(a + (r^2)/b)$, with $a = 1.0$ and $b = 0.05$.  $\Omega$ (full line), $\kappa$ (dotted line) and $\Omega \pm \frac{\kappa}{2}$ (dashed-dotted lines) are also shown.  This is the generic curve used in the tilted-ring, weak bar and weak bar$+$tilted-ring models shown in Figure~\ref{fig:quillen}, \ref{fig:barPlusquillen2} and \ref{fig:barPlusquillen3}.
\label{fig:rotation-curve}}
\end{figure}

\begin{figure}
\centering
\includegraphics[width=16cm]{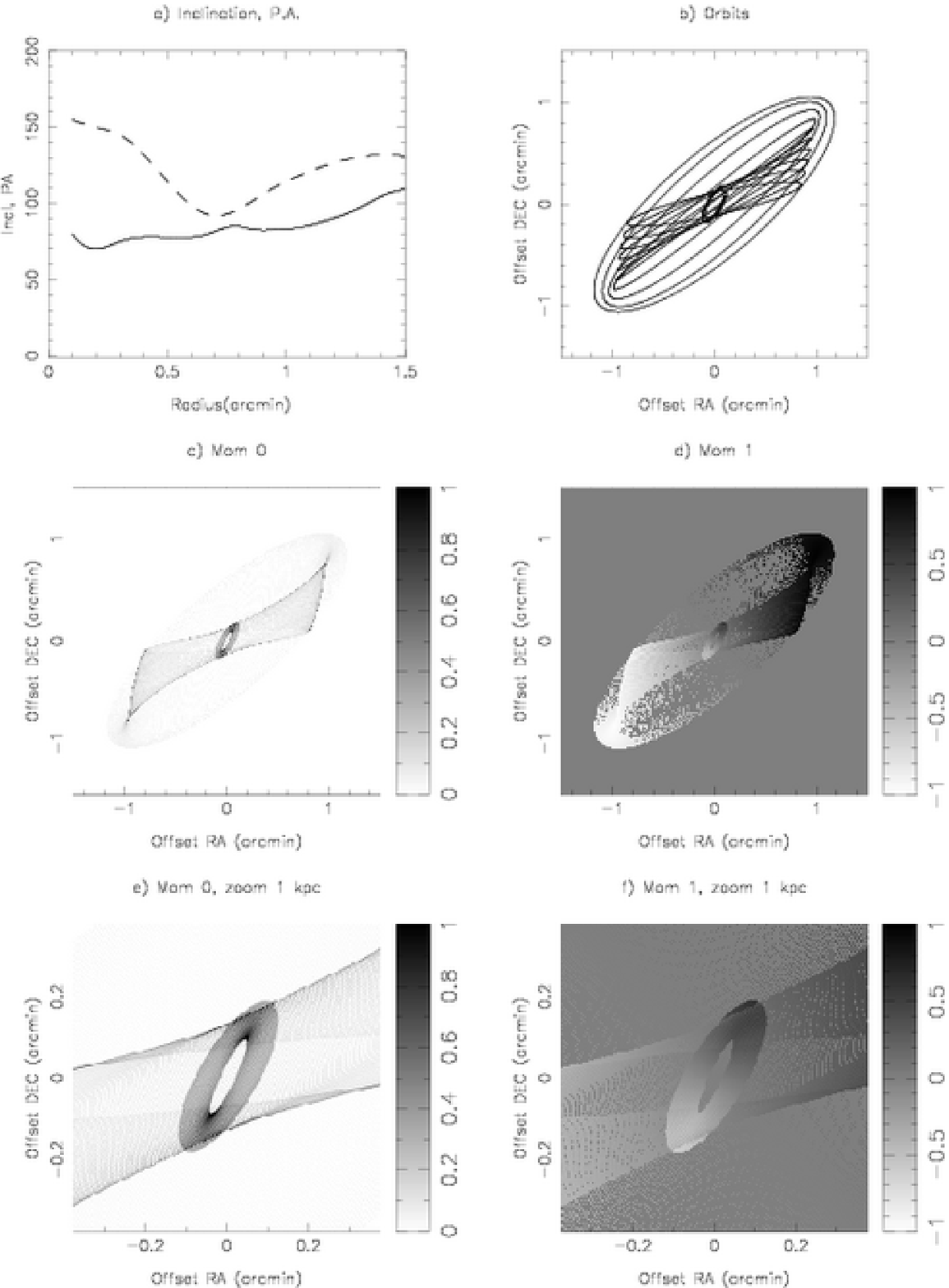}
\caption{Warped thin disk model  (\S~\ref{subsect:warp}): a) Inclination (solid line) and P.A.\ (dashed line)  as a function of radius in arcmin for the warped disk analysis, b) projected orbits, c) and d) distribution and velocity field for a field of view of 4~kpc (250\arcsec), e) and f) distribution and velocity field for a smaller field of view of 1~kpc (62\arcsec).  The intensity has been normalized so that its maximum is unity.  We use the rotation curve as shown in Figure~\ref{fig:rotation-curve}.  Inclination and P.A.\ are the same as in \citet{2006ApJ...645.1092Q}. We model a circumnuclear disk within radii r $<$ 0.2 kpc as suggested by our observations. The gap up to r = 0.8 kpc is required in this model in order to remove a bright linear feature within  the parallelogram structure (\citealt{2006ApJ...645.1092Q}, Figure 10).
\label{fig:quillen}}
\end{figure}

\begin{figure}
\centering
\includegraphics[width=12cm]{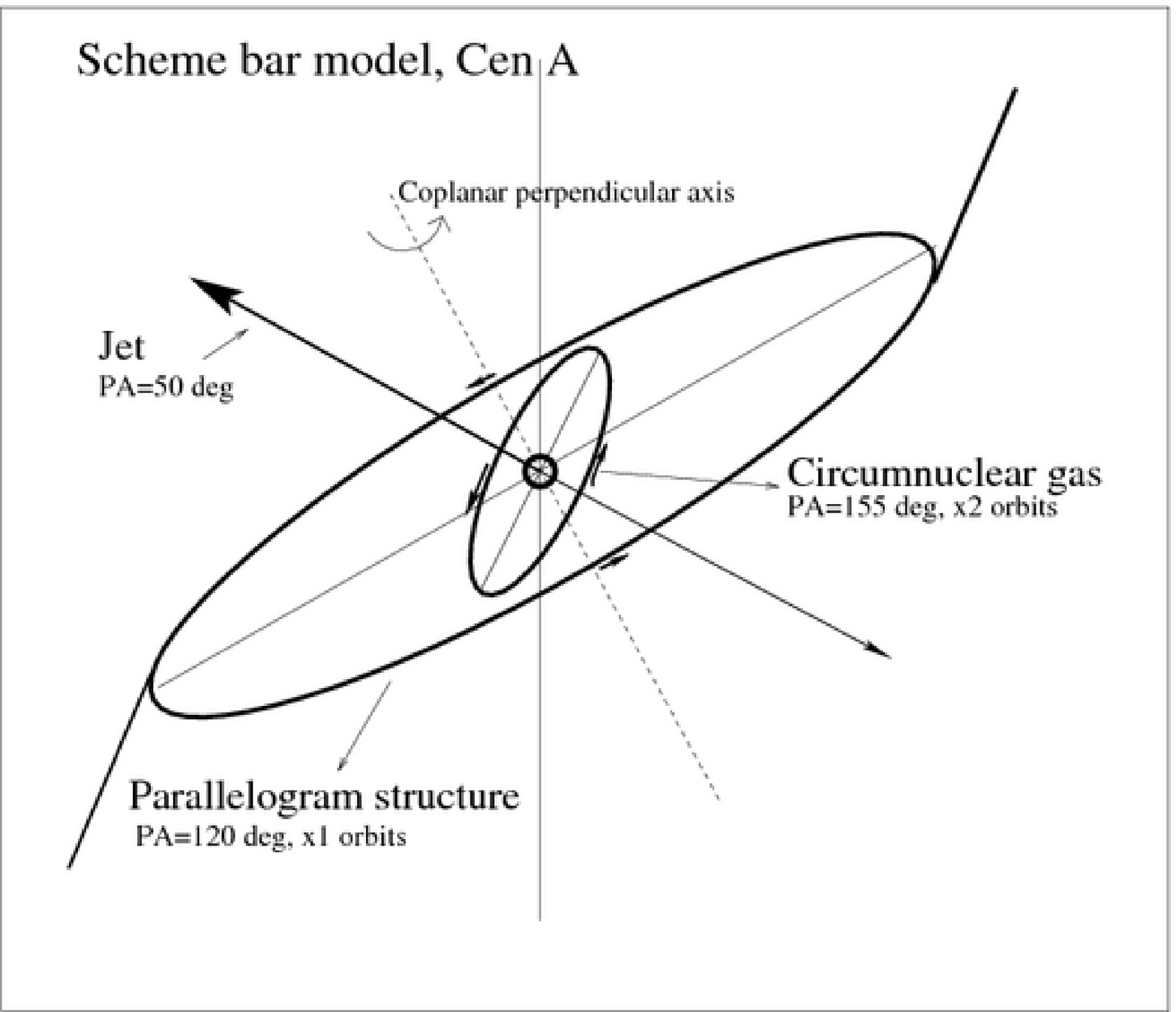}
\caption{Scheme illustrating all the components present in the inner 3~kpc of Cen~A in the proposed weak bar model scenario.}
\label{barmodelscheme}
\end{figure}

\begin{figure}
\centering
\includegraphics[width=16cm]{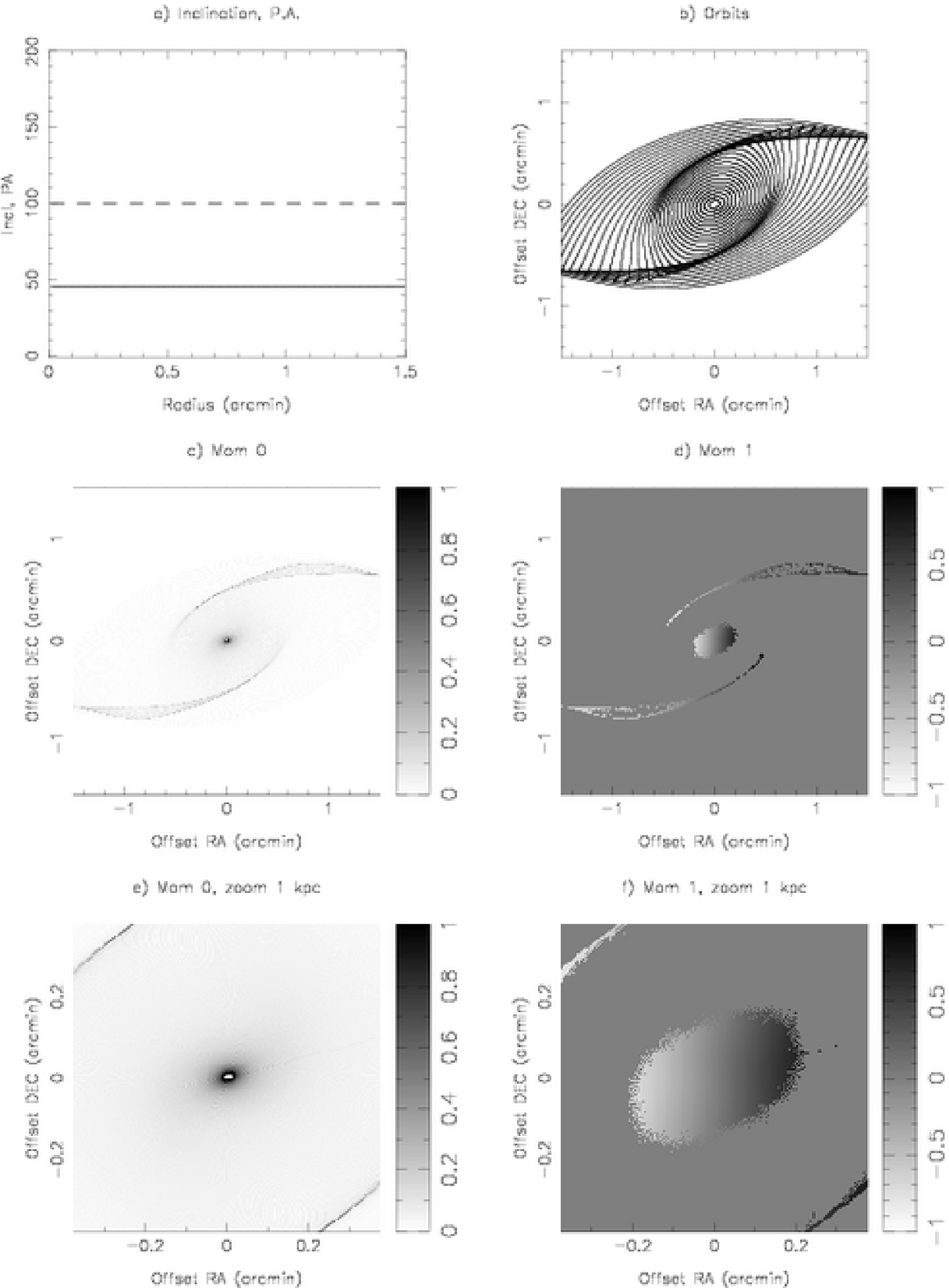}
\caption{Co-planar weak bar model  (\S~\ref{subsect:bar}). Plots are as in Figure~\ref{fig:quillen}, but here we simulate that the gas distribution is co-planar and non-circular motions using a weak bar model.  Inclination and P.A.\ are kept constant to $45\arcdeg$ and $100\arcdeg$. Note that a lower inclination to the actual i $\simeq$ $70$\arcdeg is used in order to better illustrate the appearance of the spiral arms, circumnuclear disk as well as a preferential direction with lack of emission. The model follows the formalism in \citet{1994PASJ...46..165W}. Bar potential is $\Phi_0(r) = 0.5 \times \epsilon \log(a + r^2/b)$, with $a = 1.0$ and $b=0.05$, which roughly fits the estimated rotation curve shown in Figure~\ref{fig:rotation-curve}.  We fixed $\Omega_{\rm b} = 0.3$, $\epsilon = 0.03$ and $\Lambda = 0.1$.
\label{fig:barPlusquillen2}}
\end{figure}

           
\begin{figure}
\includegraphics[width=16cm]{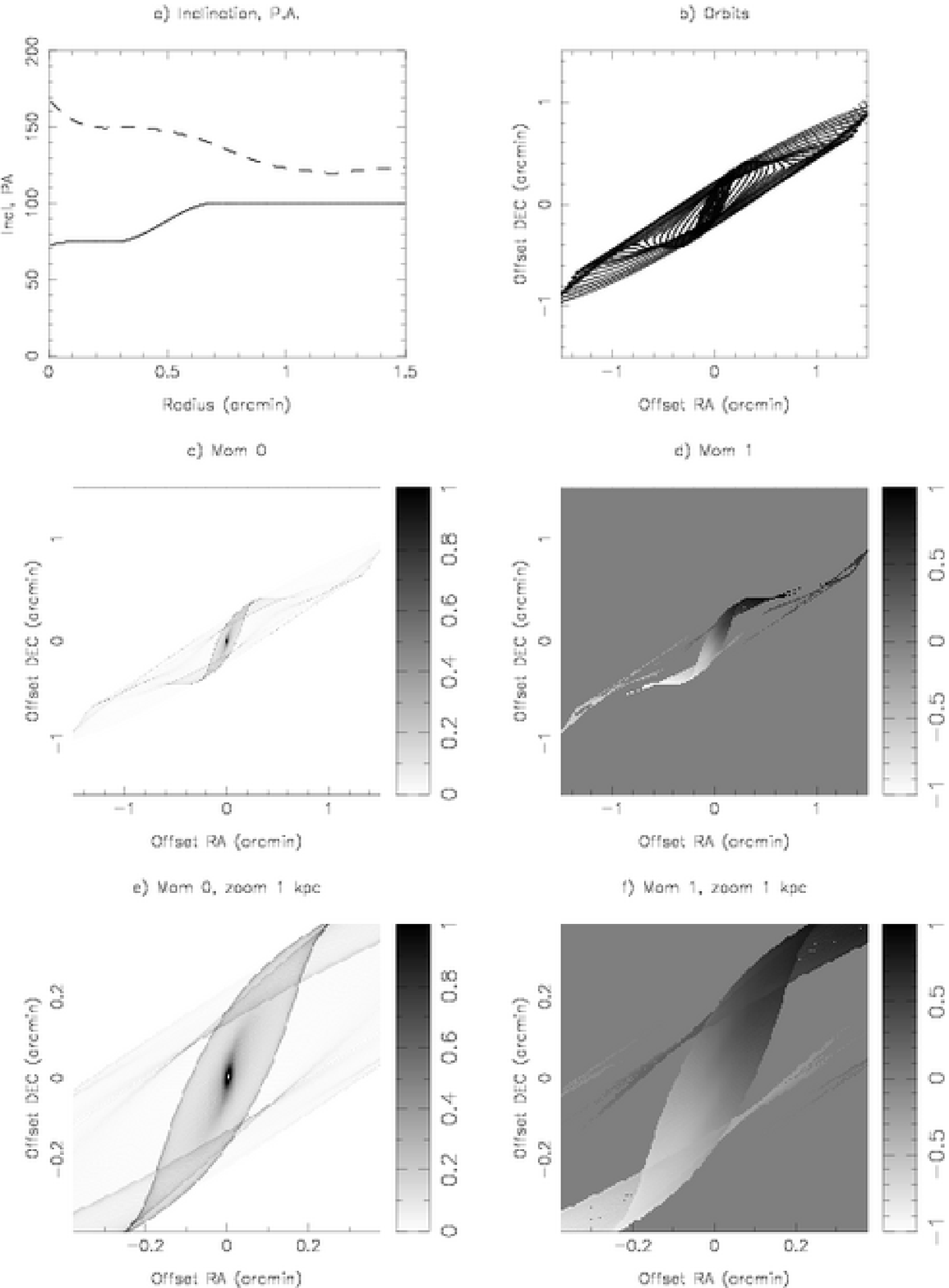}
\caption{Bar $+$ warped disk model (\S~\ref{subsect:bar+warp}): as in Figure~\ref{fig:barPlusquillen2}, we fixed $\Omega_{\rm b} = 0.3$, $\epsilon = 0.03$ and $\Lambda = 0.10$, but this time slightly varying inclination and P.A. as a function of radius. Two main constraints have been taken into account: i) P.A. from 155\arcdeg (our circumnuclear gas region) to 120\arcdeg (parallelogram filaments, dust lane); b) Inclination from i $\ge 70\arcdeg$ for the circumnuclear gas (\S~\ref{result}) with near side to the SW, to i $\simeq$ 70 -- 80\arcdeg\ with opposite near side for the gas at larger radii (indicated with values 100 -- 110\arcdeg in Fig.~\ref{fig:barPlusquillen3} a ). The rotation curve used is shown in Figure~\ref{fig:rotation-curve}.
\label{fig:barPlusquillen3}}
\end{figure}

\begin{figure}
\centering
\includegraphics[width=18cm]{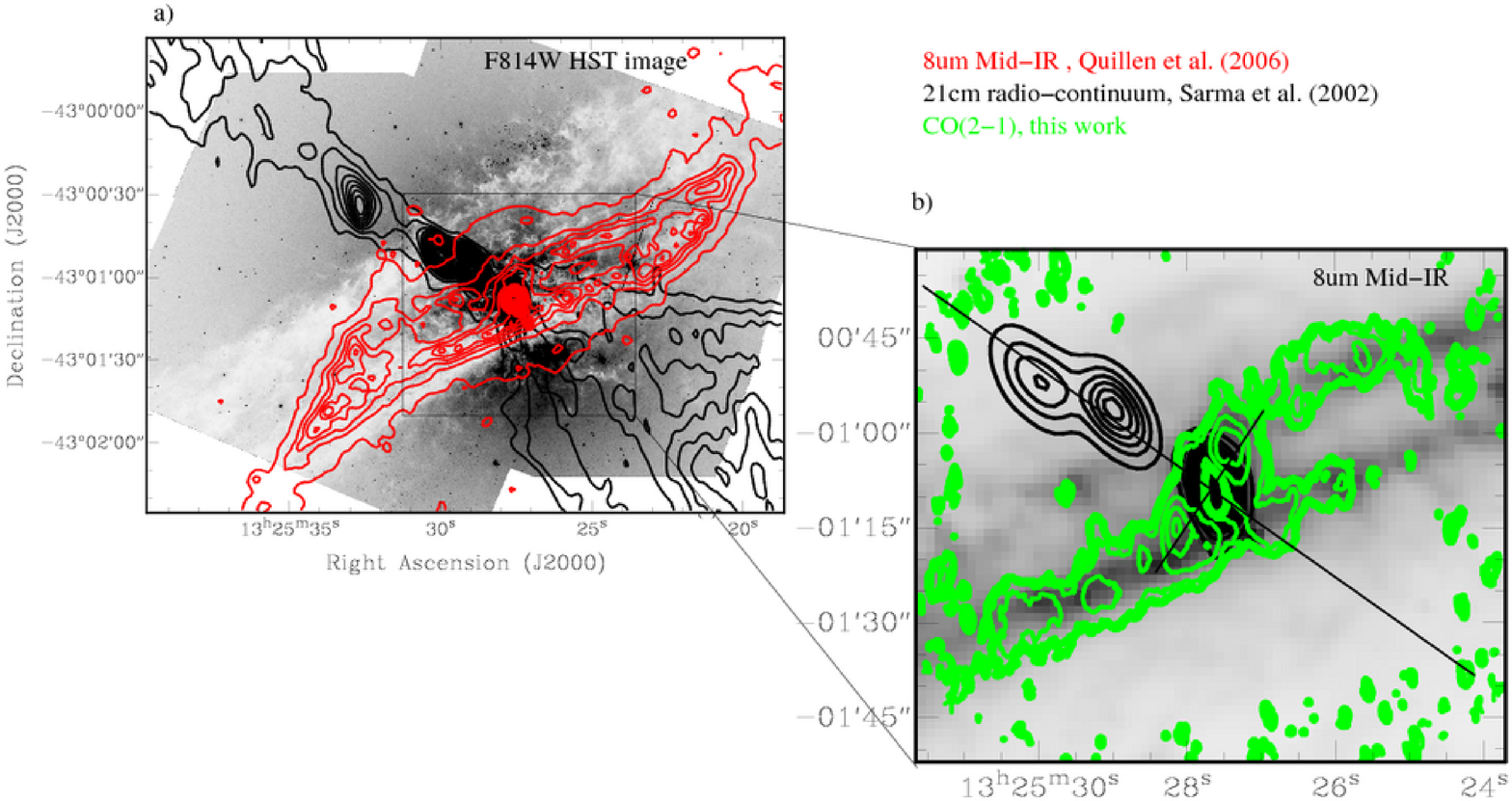}
\caption[Comparison Optical, Mom 0, mid-IR and radio-continuum]{Comparison of the dust lane location, (8 $\mu$m) dust emission distribution \citep{2006ApJ...645.1092Q},  $^{12}$CO(2--1) emission (this work) and the 21cm radio-continuum jet \citep{2002ApJ...564..696S} . 
a) HST I analog filter F814W image (\citealt{1998tx19.confE.380S}; lighter values show location of the dust lane), and superposed the Spitzer mid-IR $8~\mu$m emission in red contours (\citealt{2006ApJ...645.1092Q}; FWHM of the PSF is 2\farcs0 and the rms of the mid-IR image is 0.06~MJy~sr$^{-1}$ in steps of 30 MJy~sr$^{-1}$ ) and 21cm radio continuum emission in black contours (\citealt{2002ApJ...564..696S}; Figure~1a., nucleus and inner lobes, the beam is $8\farcs4 \times 4\farcs2$ and the rms in the image is 9~mJy~beam$^{-1}$ in steps of 90 mJy~beam$^{-1}$).  
b) Mid-IR $8~\mu$m image from Spitzer this time in grey scale (minimum is -0.06 and maximum 974.84 MJy~sr$^{-1}$).  Black contours show  \citeauthor{2002ApJ...564..696S}'s radio continuum (from 0.4 Jy beam$^{-1}$ to 6.0 in steps of 0.4 Jy beam$^{-1}$) and the green contours the  $^{12}$CO(2--1) distribution (this work, same contours as in Fig.~\ref{fig1}).  Note the clear correspondence between the main clumps of  $^{12}$CO(2--1) and mid-IR $8~\mu$m emission and that the jet is nearly perpendicular to the circumnuclear molecular gas.
\label{fig:mom0-mir}}
\end{figure}






\begin{figure}
\centering
\includegraphics[width=12cm]{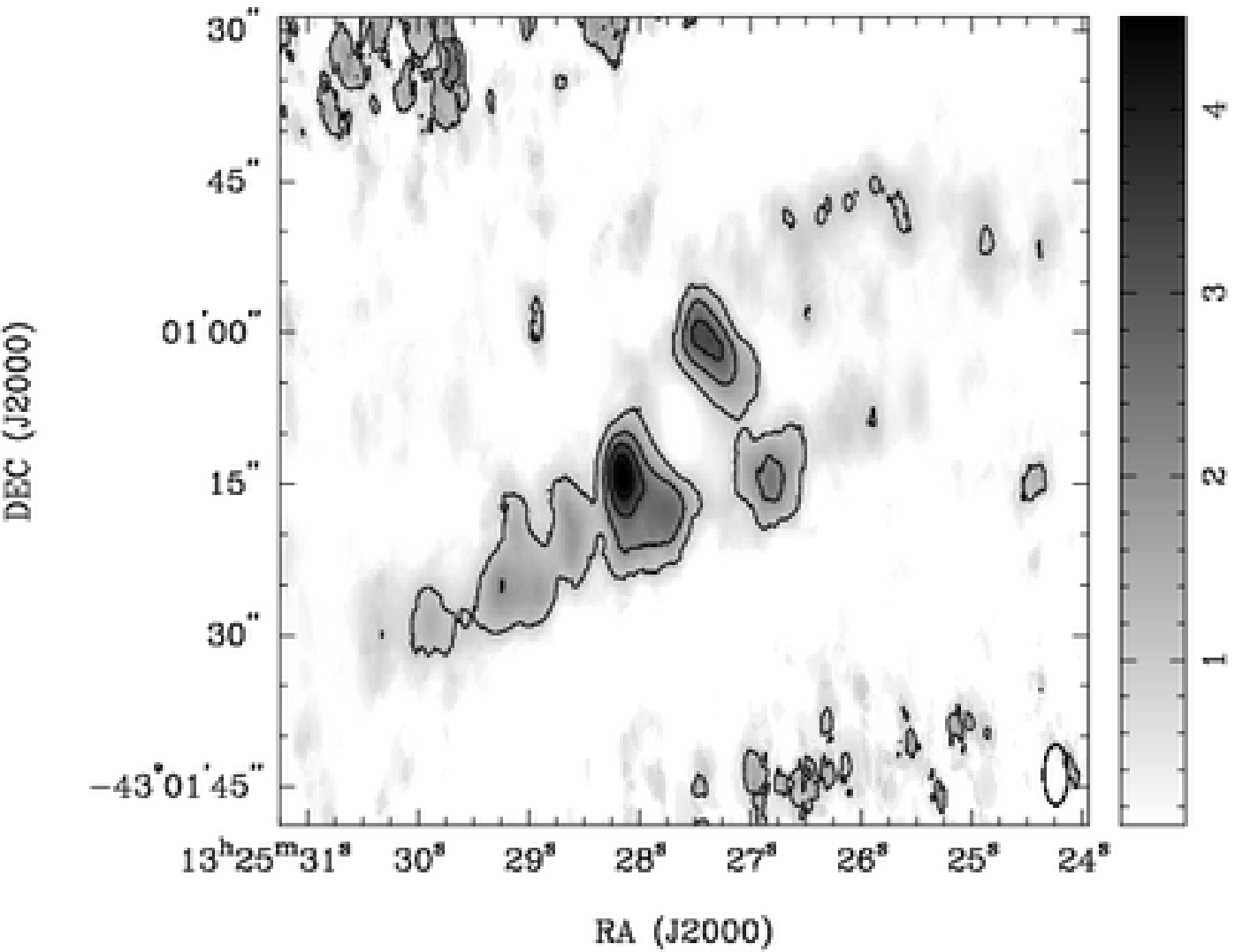}
\caption{ Ratio of integrated flux densities between the  $^{12}$CO(2--1) (as in Fig.~\ref{fig2}) and the mid-IR Spitzer $8~\mu$m  (as in Fig.\ref{fig:mom0-mir} b but convolved with a 6\farcs0 $\times$ 2\farcs4 beam), and divided by 30. Contours are at 1, 2, 3 and 4 (dimensionless since we set both in units of Jy beam$^{-1}$).
\label{fig5}}
\end{figure}



\clearpage

\begin{table}
\begin{center}
\caption{General properties of Centaurus~A (NGC~5128) \label{tbl-1}}
\begin{tabular}{ll}
\tableline
Morphology \tablenotemark{a}     & E2, peculiar \\
R.A.\ (J2000) \tablenotemark{b}  & $13^{\rm h}25^{\rm m}27\fs615$ \\
Decl.\ (J2000) \tablenotemark{b} & $-43\arcdeg01\arcmin08\farcs805$ \\
Distance (Mpc) \tablenotemark{c} & 3.42 $\pm$ 0.18 (random) $\pm$ 0.25 (systematic) \\
Linear scale ($\arcsec / {\rm pc}$)  & 16.5 \\
$D_{25}$ (\arcmin) \tablenotemark{d} & 25.7 \\
$d_{25}$ (\arcmin) \tablenotemark{d} & 20.0 \\
$V_{\rm LSR}$ (\kms)   \tablenotemark{a} & $543 \pm 2$ \\
$L_{\rm B}$, blue luminosity (L$_\odot$) \tablenotemark{e} & $3.45 \times 10^{10}$ \\ 
\mhi (L$_\odot$) \tablenotemark{a} & $(8.3 \pm 2.5) \times 10^8$ \\
$M_{\rm H_{2}}$ (M$_\odot$) \tablenotemark{e} & $3.4 \times 10^8$ \\
$L_{\rm FIR}$ (L$_\odot$) \tablenotemark{f} & $(0.6-1.8) \times 10^9$ \\
\tableline
\end{tabular}
\tablenotetext{a}{\citet{1998A&ARv...8..237I}}
\tablenotetext{b}{\citet{1998AJ....116..516M}, VLBI observations at 2.3/8.4~GHz}
\tablenotetext{c}{\citet{2007ApJ...654..186F}, distance calculated using cepheids.}
\tablenotetext{d}{\citet{1995yCat.7155....0D}}
\tablenotetext{e}{\citet{1990ApJ...363..451E}}
\tablenotetext{f}{\citet{2000A&A...359..483W}, range indicates the average value over the large scale disk as well as in the nuclear region  r $<$ 1 kpc.}
\end{center}
\end{table}

\begin{table}
\begin{center}
\caption{Main parameters of the  $^{12}$CO(2--1) SMA observations \label{tbl-2}}
\begin{tabular}{ll}
\tableline
Date           &  2006 April 5 \\
Configuration  &  7 antennas, Compact N--S \\
             &  unprojected baselines: 16 -- 70~m \\
             &  projected baselines: 6 -- 30~m \\
R.A.\ of phase center (J2000)   &  $13^{\rm h}25^{\rm m}27\fs6$ \\
Decl.\ of phase center (J2000)  &  $-43\arcdeg01\arcmin09\farcs0$ \\
Time on source (hr)       &  2 \\
FWHM of primary beam      &  52\arcsec \\
FWHM of synthesized beam  &  $2\farcs4 \times 6\farcs0$ (40 $\times$ 100~pc ), P.A.\ = $0.2\arcdeg$ \\
Velocity at band center (\kms)  &  550 \\
Total bandwidth &  2 GHz in each sideband (about 2800 \kms ) \\
              &  (separated by 10 GHz) \\
rms noise (10~\kms)             &  0.064~Jy~beam$^{-1}$ \\
Spectral resolution (\kms)      &  1~\kms \\
\tableline
\end{tabular}
\end{center}
\end{table}

\begin{table}
\caption{Derived parameters of the  $^{12}$CO(2--1) emission lines \label{tbl-3}}
\begin{center}
\begin{tabular}{l l l l r r}
\tableline
\small Component   &\small  Peak Flux Density &\small Mean Vel. &\small Vel. Width &\small $S_{\rm  ^{12}CO(2-1)}$ &\small      $M_{\rm H_{2}}$ \\
          &\small (Jy~beam$^{-1}$ km s$^{-1}$)  &\small (km s$^{-1}$) &\small  (km~s$^{-1}$) &\small  (Jy~km~s$^{-1}$) &  \small     (M$_{\odot}$) \\
\tableline
Broad line  &      126 $\pm$ 1  &  560 $\pm$ 15 &  314 $\pm$ 15  &   1182 $\pm$  5   & $(0.6\pm0.1)\times10^7$ \\
(Circumnuclear gas)  & & & & & \\
Narrow line &                   &  535 $\pm$ 15 &  160 $\pm$ 15  &   2324 $\pm$ 14   & $(5.9\pm0.3)\times10^7$ \\
(Outer gas) & & & & &  \\
All         &                   &               &                &   3506 $\pm$ 15   & $(6.5\pm0.3)\times10^7$ \\
\tableline
\end{tabular}
\tablecomments{We adopt a galactic CO-to-H$_2$ conversion factor  $X = 0.4 \times 10^{20}$~cm$^{-2}$~(K~\kms)$^{-1}$ for the circumnuclear gas and  $X = 2.0 \times 10^{20}$~cm$^{-2}$~(K~\kms)$^{-1}$ for the outer gas, estimated following recent values for X for our own galaxy as a function of radius \citep{2001ApJ...547..792D}.  We use equations in \S.~4 of \citet{1999ApJS..124..403S} for molecular gas mass $M_{\rm H_{2}}$ and gas surface density $\Sigma_{\rm gas}$ using $M_{\rm gas}$ = 1.36 $M_{\rm H_{2}}$, where the factor 1.36 accounts for elements other than hydrogen \citep{1973asqu.book.....A}.}
\end{center}
\end{table}

\begin{table}
\caption{Line of sight velocity curve \label{tab-alpha}}
\begin{center}
\begin{tabular}{lccrl}
\tableline
Position range & Velocity gradient ($dv/dr$) & Resolution     & Line & Comment \\
(pc (\arcsec)) &   (km~s$^{-1}$~pc$^{-1}$)   & (\arcsec)      &      & \\
\tableline
80 --  165  (5 -- 10)  & 1.2               & $6.0\times2.4$ &  $^{12}$CO(2--1)\tablenotemark{a}   & Circumnuclear gas \\
80 --  335  (5 -- 20)  & 0.2               & $6.0\times2.4$ &  $^{12}$CO(2--1)\tablenotemark{a}   & Outer molecular gas \\
30 -- 3000  (2 -- 180) & 0.3               & $2\times2$     & H$\alpha$\tablenotemark{b}  & Up to 1~kpc, then flat \\
750 -- 2500 (45 -- 150) & 0.2               & $45\times45$   &  $^{12}$CO(1--0)\tablenotemark{c}   & Up to 1.5~kpc \\
235 --  835 (14 -- 50)  & 0.2               & $14\times14$   &  $^{12}$CO(3--2)\tablenotemark{d}   & Outer molecular gas \\
1000 -- 5000 (60 -- 300) & $\simeq$0.1       & $77\times33$   & \ion{H}{1}\tablenotemark{e} & Up to 2~kpc \\
\tableline
\end{tabular}
\end{center}
\tablenotetext{a}{This work}
\tablenotetext{b}{\citet{1992ApJ...387..503N}}
\tablenotetext{c}{\citet{1990ApJ...363..451E}}
\tablenotetext{d}{\citet{2001AA...371..865L}}
\tablenotetext{e}{\citet{1990AJ.....99.1781V}}
\end{table}

\end{document}